\documentclass[12pt]{article}
\usepackage{amsmath,amssymb,multicol,epsf,rotate}
\renewcommand{\theequation}{\thesection.\arabic{equation}}

\begin{document}

\title{Wall Induced Density Profiles and Density \\
  Correlations in Confined Takahashi Lattice Gases}

\author{J.~Buschle, P.~Maass and W.~Dieterich\\[0.5cm]
\it Fakult\"at f\"ur Physik, Universit\"at Konstanz,\\[0.1cm]
\it D-78457 Konstanz, Germany}
 
\date{August 17, 1999\\[0.3cm](revised November 3, 1999)}

\maketitle

\begin{abstract}
  
  We propose a general formalism to study the static properties of a
  system composed of particles with nearest neighbor interactions that
  are located on the sites of a one-dimensional lattice confined
  by walls (``confined Takahashi lattice gas''). Linear recursion
  relations for generalized partition functions are derived, from
  which thermodynamic quantities, as well as density distributions and
  correlation functions of arbitrary order can be determined in the
  presence of an external potential. Explicit results for density
  profiles and pair correlations near a wall are presented for various
  situations. As a special case of the Takahashi model we consider in
  particular the hard rod lattice gas, for which a system of nonlinear
  coupled difference equations for the occupation probabilities has
  been presented previously by Robledo and Varea. A solution of these
  equations is given in terms of the solution of a system of
  independent linear equations. Moreover, for zero external potential
  in the hard rod system we specify various central regions between
  the confining walls, where the occupation probabilities are constant
  and the correlation functions are translationally invariant in the
  canonical ensemble.  In the grand canonical ensemble such regions do
  not exist.

\end{abstract}

\section{Introduction}\label{sec_introduction}
The understanding of the static and dynamic behavior of fluids in
confined geometries is a problem of active current research
\cite{Croxton:1986,Chavrolin/etal:1986}. This research is largely
motivated by technological applications where one wants to create
small surface structures with suitable physical and chemical
properties \cite{Delamarche:1996}. The question, how the formation of
such structures is influenced by confining walls, has raised the
interest in many basic phenomena, such as the development of density
profiles, pair correlations and ordering effects at surfaces
\cite{Binder:1986}, wetting transitions
\cite{Dieterich:1988,Schick:1990}, or a variety of surface induced
kinetic processes \cite{Puri/Frisch:1997}. In general, an exact
analytical treatment of the various effects is not possible and one
has to rely on approximation schemes. Well established techniques for
this purpose are the density functional theory with its many variants
(see \cite{Evans:1992} and references therein), the cluster variation
\cite{Kikuchi:1951} and path probability method \cite{Kikuchi:1966},
as well as the classical thermodynamic perturbation theories (see
e.g.~\cite{Salter/Davies:1975}).

In one-dimensional fluid systems, however, exact results can be obtained for
special models. Such results have their merits in establishing ``boundary
conditions'' for the development of approximate theories in higher dimensions
$d>1$ in the sense that these should become exact for $d=1$.  Physically, this
issue in particular pertains to theories for confined systems with tunable
confinement, allowing a ``dimensional crossover'' form the case $d>1$ to $d=1$
\cite{Rosenfeld/etal:1996}.  Exact findings also provide the possibility to
systematically test the quality of approximations. Moreover, for many
phenomena, such as e.g. the emergence of the well-known density oscillations
of fluids near hard walls, the dimensionality does not seem to play a crucial
role, and valuable insight into the origin of these phenomena may be gained by
investigating appropriate one-dimensional reference systems.

The first exact density functional in $d=1$ was set up by Percus
\cite{Percus:1976} for a fluid of hard rods. This functional yields an
integral equation for the density profile in an arbitrary external potential.
Later Percus showed that a further exact density functional can be written
down for the special ``sticky core'' model \cite{Percus:1982} that, in
addition to the hard-rod repulsion, includes a ``zero-range'' attractive force
between nearest neighbor rods. This was subsequently generalized to
finite-range forces between neighboring rods \cite{Percus:1989}.
A generalized discrete version of the continuum hard-rod fluid on a
linear chain was studied by Robledo and Varea
\cite{Robledo/Varea:1981}. They derived an exact functional for the
mean occupation numbers of the rod centers on the chain, which, by
taking the continuum limit, allowed them to recover the continuum
density functional of Percus (for a review on classical density
functionals, see \cite{Percus:1994}). Within this theory, the discrete
hard-rod model leads to a rather complicated system of nonlinear
finite difference equations for the mean occupation numbers in an
arbitrary external potential, whose numerical solution requires a
considerable computational effort.

In this article we will show that a more general discrete
one-dimensional system can be considered, which allows one to
calculate bulk and surface thermodynamical properties as well as
equilibrium density profiles and density correlations in arbitrary
external potentials. The system is an extension of the continuum
Takahashi model \cite{Takahashi:1942} to a lattice gas model, in which
only neighboring particles interact with each other.  We first show
that the canonical and grand canonical partition functions of this
``Takahashi lattice gas'' (TLG) obey simple recursion relations and
that density profiles and density correlations can be conveniently
calculated from the partition functions due to the one-dimensional
nature of the model. Since the TLG contains as a special case the
hard-rod model studied by Robledo and Varea \cite{Robledo/Varea:1981}
(for which the interaction potential would be infinite for
interparticle distances smaller than the rod length and zero else), we
can give a simple solution of the nonlinear difference equations
derived in \cite{Robledo/Varea:1981}. In this context we found it
worthwhile to rederive the central formulae given in
\cite{Robledo/Varea:1981}.  We will apply our formalism to a system of
hard-rods confined both by hard and soft walls, and also to a system
of particles, which in addition to the athermal hard-rod repulsion
experience a finite interaction potential over a limited range. For
these different cases density profiles and density correlations near
the confining walls will be discussed in detail and compared with each
other.

\section{Takahashi Lattice Gas}\label{takahashi-sec}
\setcounter{equation}{0}

In the TLG we consider $N$ particles at positions $i_k$,
$k=1,\ldots,N$, on a linear chain with $M$ sites $i=1,\ldots,M$. No
more than one particle is allowed to occupy a given lattice site. Two
neighboring particles separated by $n-1$ vacant lattice sites interact
via a potential $v(n)$. There is no interaction between particles that
are not nearest neighbors, that means between particles that have at
least one other particle in between them. In addition the particles
experience an external potential $u(i)$ and it is assumed that two
confining walls are present at the boundary sites $i=0$ and $i=M+1$.
These are modeled by two additional particles that are held fixed at
the boundary sites.  The energy of a particle configuration $1\le
i_1<\ldots<i_N\le M$ is then given by
\begin{equation}
\beta H=\sum_{k=1}^{N} u(i_k)+
v(i_1)+\sum_{k=2}^{N} v(i_k-i_{k-1})+ v(M+1-i_N)\,,
\label{h-eq}
\end{equation}
where $\beta=1/k_BT$ and we have assumed for simplicity that the particles at
the boundary sites are the same as those on the chain. It will become clear in
the following that one could also consider a modified interaction with the
walls.

\subsection{General case of arbitrary external potential}
\label{gencase-subsec}
In the presence of a (non-constant) external potential $u(i)$,
it is convenient to define a generalized canonical partition function by
\begin{equation}
Z(N,M',\alpha)=\hspace{-0.5cm}\sum_{1\le i_1<\ldots i_N\le M'}
\hspace{-0.3cm}\exp-\left[\sum_{k=1}^{N} u(i_k+\alpha)+
v(i_1)+\sum_{k=2}^N v(i_k-i_{k-1})+ v(M'+1-i_N)\right]\,,
\label{z-eq}
\end{equation}
for integers $\alpha\ge0$ and $M'+\alpha\le M$. Equation~(\ref{z-eq})
defines $Z(N,M',\alpha)$ if $N\le M'$, while for $N>M'$ we set
$Z(N,M',\alpha)\equiv0$. Note that for $\alpha=0$ and $M'=M$ we
recover the ordinary partition function $Z(N,M)\equiv Z(N,M,0)$. As
far as only thermodynamic quantities shall be calculated, we could
limit ourselves to the conventional form, but to evaluate density
profiles and density correlations for the TLG, we need to consider the
generalized functions (see below). Separating the summation over the
positions $l=i_N$ of the rightmost particle, we can write
\begin{eqnarray}
Z(N,M',\alpha)&=&\sum_{l=N}^{M'}
\exp\left[-v(M'+1-l)-u(l+\alpha)\right]\times\nonumber\\
&&\sum_{1\le i_1<\ldots<i_{N-1}\le l-1}\exp-\Bigl[
\sum_{k=1}^{N-1}u(i_k+\alpha)+
v(i_1)+\sum_{k=2}^{N-1} v(i_k-i_{k-1})+ v(l-i_{N-1})
\Bigr]\,\nonumber\\
&=&\sum_{l=1}^{M'}
\exp\left[-v(M'+1-l)-u(l+\alpha)\right] Z(N-1,l-1,\alpha)\,.
\label{recurz-eq}
\end{eqnarray}
In the last line we could start the summation from $l=1$ because
of our setting $Z(N,M',\alpha)\equiv0$ for $N>M'$. The recursion
relation (\ref{recurz-eq}) is so far valid for $N>1$. It becomes valid also
for $N=1$ if we set $Z(0,M',\alpha)\equiv\exp[-v(M'+1)]$.

From eq.~(\ref{recurz-eq}) we readily derive a recursion relation for the
analogous generalized grand partition function,
\begin{eqnarray}
\phantom{.}\hspace{-1cm}\Omega(\lambda,M',\alpha)\!
&=&\!\sum_{N=0}^\infty Z(N,M',\alpha)\lambda^N\\
&=&\exp\left[-v(M'+1)\right]+\lambda\sum_{l=1}^{M'}
\exp\left[-u(l+\alpha)-v(M'+1-l)\right]\Omega(\lambda,l-1,\alpha)\nonumber
\label{recurom-eq}
\end{eqnarray}
As before, we define $\Omega(\lambda,M)\equiv \Omega(\lambda,M,0)$.
This recursion relation (\ref{recurom-eq}) is valid for $M'>1$, but can be
made valid also for $M'=1$ if we set
$\Omega(\lambda,0,\alpha)\equiv\exp[-v(1)]$.  The
fugacity $\lambda$ determines, for given $M$, the mean number $N(\lambda,M)$ of
particles in the system.

Using the recursion relations (\ref{recurz-eq}, \ref{recurom-eq}) one may
calculate all thermodynamic properties for a given interaction $v(n)$ and
external potential $u(i)$ by taking $\alpha=0$ and $M'=M$. We now show how one
can calculate also density profiles and correlations (of arbitrary order),
once $Z(N,M',\alpha)$ or $\Omega(\lambda,M',\alpha)$ has been calculated from
eqs.~(\ref{recurz-eq},\ref{recurom-eq}). To this end we first determine the
probability $w(l,r)$ to find the $r$th particle at position $i_r=l$,
\begin{eqnarray}
w(l,r)
&=&Z^{-1}(N,M)\hspace{-0.7cm}\sum_{\ldots<i_{r-1}<l<i_{r+1}<\ldots}
\hspace{-0.7cm}\exp-\Bigl[
  v(i_1)+\sum_{k=2}^N v(i_k-i_{k-1})+ v(M+1-i_N)+\sum_{k=1}^{N} u(i_k)\Bigr]
             \nonumber\\[0.7cm]
&=&\exp[-u(l)]\,Z^{-1}(N,M)\nonumber\\
   &&\hspace{0.15cm}{}\times\hspace{-0.3cm}
     \sum_{1\le i_1<\ldots<i_{r-1}\le l-1}\hspace{-0.7cm}
     \exp-\Bigl[v(i_1)+\sum_{k=2}^{r-1} v(i_k-i_{k-1})+ v(l-1+1-i_{r-1})
     +\sum_{k=1}^{r-1} u(i_k)\Bigr]\nonumber\\
  &&\hspace{0.15cm}{}\times\hspace{-0.3cm}
     \sum_{l+1\le i_{r+1}<\ldots<i_N\le M}
    \hspace{-0.7cm}\exp-\Bigl[v(i_{r+1}-l)+
     \sum_{k=r+2}^{N} v(i_k-i_{k-1})+ v(M+1-i_N)
      +\sum_{k=r+1}^{N} u(i_k)\Bigr]\nonumber\\[0.7cm]
&=&\exp[-u(l)]\,Z^{-1}(N,M)\,Z(r-1,l-1)\nonumber\\
  &&\hspace{0.15cm}{}\times\hspace{-0.4cm}
     \sum_{1\le j_1<\ldots<j_{N-r}\le M-l}\hspace{-0.8cm}
     \exp-\Bigl[v(j_1)+\sum_{k=2}^{N-r} v(j_k-j_{k-1})+v(M-l+1-j_{N-r})+
     \sum_{k=1}^{N-r} u(j_k+l)\Bigr]\nonumber\\[0.7cm]
&=&\exp[-u(l)]\,\frac{Z(r-1,l-1)Z(N-r,M-l,l)}{Z(N,M)}\,.
\label{wlr-eq}
\end{eqnarray}
In the second step we have introduced the shifted particle positions
$j_s=i_{r+s}-l$ for $s=1,\ldots,N-r$. By doing this, $u(i_k)$ transforms to
$u(j_k+l)$ and it becomes clear now why we had to introduce the generalized
partition function $Z(N,M',\alpha)$ with $\alpha\ne0$. In the canonical
ensemble the probability $p(l;N,M)$ for the site $l$ to be occupied is then
given by
\begin{equation}
p(l;N,M)=\sum_{r=1}^N w(l,r)=\frac{\exp[-u(l)]}{Z(N,M)}
\sum_{r=1}^N Z(r-1,l-1)\,Z(N-r,M-l,l)\,.
\label{plnm-eq}
\end{equation}
In the grand canonical ensemble the convolution in eq.~(\ref{plnm-eq})
factorizes and we obtain the corresponding occupation probability
\begin{equation}
\tilde p(l;\lambda,M)=\lambda\exp[-u(l)]\,
\frac{\Omega(\lambda,l-1)\,\Omega(\lambda,M-l,l)}{\Omega(\lambda,M)}\,.
\label{pllm-eq}
\end{equation}
We like to note that for a symmetric external potential,
$u(i)=u(M+1-i)$, it follows that
$\Omega(\lambda,M-l,l)=\Omega(\lambda,M-l)$. Hence it suffices to
calculate $\Omega(\lambda,M',\alpha)$ from eq.~(\ref{recurom-eq}) for
$\alpha=0$ to obtain the density profile in this symmetric case.

By an analogous decomposition of the partition function into
products of generalized partition functions (corresponding to
various system sizes) one can derive the joint probabilities
$p_s(l_1,\ldots,l_s;N,M)$ to find the sites $l_1<\ldots<l_s$
being occupied in the canonical ensemble,
\begin{eqnarray}
p_s(l_1,\ldots,l_s;N,M)\!&=&\!\frac{\exp\left[-\sum_{k=1}^s u(l_k)\right]}
{Z(N,M)}
\hspace{-0.6cm}\sum_{1\le r_1<\ldots< r_s\le N}\hspace{-0.6cm}
\Bigl\{Z(r_1-1,l_1-1)\,Z(N-r_s,M-l_s,l_s)\nonumber\\
&&\hspace{3cm}\times\prod_{k=1}^{s-1} 
Z(r_{k+1}-r_k-1,l_{k+1}-l_k-1,l_k)\Bigr\}\,.
\label{ps-eq}
\end{eqnarray}
From this we obtain the corresponding joint probabilities in the grand
canonical ensemble,
\begin{eqnarray}
\tilde p_s(l_1,\ldots,l_s;\lambda,M)\!&=&\!
\frac{\lambda^s\exp\left[-\sum_{k=1}^s u(l_k)\right]}{\Omega(\lambda,M)}
\,\Omega(\lambda,l_1-1)\,\Omega(\lambda,M-l_s,l_s)\nonumber\\
&&\hspace{3cm}\times\prod_{k=1}^{s-1}\Omega(\lambda,l_{k+1}-l_k-1,l_k)\,.
\label{psl-eq}
\end{eqnarray}
Note that for $s=1$ eqs.~(\ref{ps-eq},
\ref{psl-eq}) reduce to eqs.~(\ref{plnm-eq}, \ref{pllm-eq}). From
eqs.~(\ref{recurz-eq}, \ref{ps-eq}) or eqs.~(\ref{recurom-eq}, \ref{ps-eq}) one
can readily calculate density profiles and density correlations of arbitrary
order in the canonical or grand canonical ensemble for arbitrary interaction
$v(n)$ and external potential $u(i)$.

\subsection{Special case of vanishing external potential}
\label{specase}
In case of a vanishing (or constant) external potential it is not
needed to introduce the generalized partition functions in
eqs.~(\ref{z-eq}, \ref{recurom-eq}) and accordingly one can set the
third argument $\alpha$ in $\Omega(\lambda,M,\alpha)$ equal to zero in
all formulae in Sec.~\ref{gencase-subsec}. The occupation
probabilities $p(l;N,M)$ and $\tilde p(l;\lambda,M)$ in
eqs.~(\ref{plnm-eq},\ref{pllm-eq}) can be written as
\begin{equation}
p(l;N,M)=
\sum_{r=1}^N Z(r-1,l-1)\,\frac{Z(N-r,M-l)}{Z(N,M)}\,,\hspace{0.3cm}
\tilde p(l;\lambda,M)=\lambda\frac{\Omega(\lambda,l-1)\Omega(\lambda,M-l)}
{\Omega(\lambda,M)}\,,
\label{pllmsimp-eq}
\end{equation} 
and analogous simplifications are obtained for the joint probabilities of
higher order in eq.~(\ref{psl-eq}).

Moreover, we can solve the recursion relations
(\ref{recurz-eq},\ref{recurom-eq}) explicitly in terms of the
generating functions $H(N,s)=\sum_{M=0}^{\infty} Z(N,M)s^M$ and
$G(\lambda,s)=\sum_{M=0}^{\infty} \Omega(\lambda,M)s^M$, which are
explicitly given by
\begin{equation}
H(N,s)=\frac{\varphi(s)^{N+1}}{s}\,,\hspace{0.5cm}
G(\lambda,s)=\frac{\varphi(s)}{s[1-\lambda\varphi(s)]} 
\label{gls-eq}
\end{equation}
with 
\begin{equation}
\varphi(s)=\sum_{l=1}^{\infty} \exp[-v(l)] s^l\,.
\label{varphi-eq}
\end{equation}

If $v(l)$ has a finite range, that means $v(l)=0$ for $l\ge l_0$, we obtain
from (\ref{gls-eq}) $G(\lambda,s)=P(\lambda,s)/Q(\lambda,s)$, where
$P(\lambda,s)$ and $Q(\lambda,s)$ are polynomials in $s$ of degree $l_0-1$ and
$l_0$, respectively.  According to a theorem for rational generating functions
\cite{Stanley:1986}, $\Omega(\lambda,M)$ then has the form
$\Omega(\lambda,M)=\sum_{j=0}^k c_j(\lambda,M) s_j(\lambda)^{-M}$, where
$s_j(\lambda)$, $j=0,\ldots,k$ are the distinct zeros of $Q(\lambda,s)$ with
multiplicities $d_j$, and $c_j(\lambda,M)$ are polynomials in $M$ of degree
less than $d_j$. The moduli of the zeros $s_j$ are considered to be ordered,
$|s_0|\le |s_1|\le \ldots \le |s_k|$.

As shown in Appendix A, $s_0$ is real with $0<s_0<1$, $d_0=1$, and $|s_j|>s_0$
for $j=1,\ldots,k$. Hence we can write
$\Omega(\lambda,M)=s_0^{-M}[c_0+\sum_{j=1}^k c_j(M) (s_0/s_j)^M]$ and obtain
\begin{equation}
\Omega(\lambda,M)\sim c_0(\lambda) s_0(\lambda)^{-M}
\end{equation}
in the thermodynamic limit $M\to\infty$.  The one-to-one
correspondence between the fugacity $\lambda$ and the number density
$\bar p=N(\lambda,M)/M$ in this limit follows from the relations
(see Corollary A.1. in Appendix A)
\begin{equation}
\bar p=\varphi(s_0)/[s_0\varphi'(s_0)]\,,\hspace{0.5cm}
\lambda=1/\varphi(s_0)\,.
\label{barp-eq}
\end{equation}
Using the asymptotic limit for $\Omega(\lambda,M)$ we obtain from
eq.~(\ref{pllmsimp-eq})
\begin{equation}
p_\infty(l;\lambda)\equiv\lim_{M\to\infty}
\tilde p(l;\lambda,M)=\lambda\Omega(\lambda,l-1)s_0^l\,.
\label{pinf-eq}
\end{equation}
In fact, as shown in Appendix A, eqs.~(\ref{barp-eq}, \ref{pinf-eq})
hold true even for a more general interaction potential $v(l)$, which
for $l$ larger than some $l_\star$ is bounded and for $l\to\infty$
approaches zero.  The analogous occupation probability
$p_\infty(l)$ in the canonical ensemble is the same as 
$p_\infty(l;\lambda)$, if for given $\bar p$ the corresponding unique
fugacity $\lambda$ is used (see eq.~(\ref{barp-eq}) and Appendix A).
Moreover, the joint probabilities $\tilde p_s(l_1,\ldots,l_s;\lambda,M)$
in the grand-canonical ensemble (and the corresponding
$p_s(l_1,\ldots,l_s;N,M)$ in the canonical ensemble) factorize
in terms of $p_\infty(l;\lambda)$ in the thermodynamic limit, that means
\begin{equation}
p_{s,\infty}(l_1,\ldots,l_s;\lambda)\equiv 
\lim_{M\to\infty}\tilde p_s(l_1,\ldots,l_s;\lambda,M)=
  p_\infty(l_1;\lambda)\prod_{k=2}^s p_\infty(l_k-l_{k-1};\lambda)\,.
\end{equation}
For the special case of a finite range interaction potential
considered above (i.e. $v(l)=0$ for $l\ge l_0$) there exists a
constant $C>0$ such that $|p_\infty(l;\lambda)-\bar p|<C l^\nu
e^{-l/\xi}$, where $\nu\le l_0-2$ is an integer, and $\xi=-1/\ln(r)$
with $r=\max_{1\le j\le k}\{s_0/|s_j|\}<1$.

\section{Hard-Rod Lattice Gas Revisited}\label{hardrod-sec}
\setcounter{equation}{0}

A particularly simple situation occurs, when the interaction potential in
eq.~(\ref{h-eq}) is given by
\begin{equation}
v(n)=v_{\scriptscriptstyle\rm HR}(n)
\equiv\left\{\begin{array}{r@{\hspace{1cm}}l}\infty\,, & 0\le n<2m \\[0.2cm]
                            0\,, & n\ge2m\end{array}\right.
\label{hardrodpot-eq}
\end{equation}
This potential can be viewed as describing a system composed of hard-rods with
lengths $2m$ (with hard walls at positions $m$ and $M+1-m$ due to the fixed
hard-rods at positions $0$ and $M+1$). 

Setting the mass density of the rods equal to unity, we can express
the local mass density $p_{\rm mass}(l;N,M)$ along the one-dimensional
chain by the occupation probabilities $p(l;N,M)$ (that refer to the rod
centers) according to
\begin{equation}
p_{\rm mass}(l;N,M)=\frac{1}{2}[p(l-m;N,M)+p(l+m;N,M)]+\sum_{j=-(m-1)}^{m-1} 
p(l-j;N,M)\,.
\label{rho-eq}
\end{equation}
This formula holds true in the canonical as well as in the grand canonical
ensemble (if $p_{\rm mass}(l;N,M)$ is replaced by $\tilde p_{\rm
  mass}(l;\lambda,M)$ and $p(l;N,M)$ by $\tilde p(l;\lambda,M)$). Note that
the total number $N$ of rods must be smaller than $M/2m$.

\subsection{Explicit results for homogeneous systems}

Density profiles and density correlations can be calculated explicitly
in the absence of an external potential by using the general method
developed in Sec.~\ref{specase}.  From eq.~(\ref{gls-eq}) we find
$\varphi(s)=s^{2m}/(1-s)$ and $G(\lambda,s)=s^{2m-1}/(1-s-\lambda
s^{2m})$, and therefore
\begin{equation}
\Omega(\lambda,l)=\sum_{n=0}^\infty {l-(2m-1)(n+1)\choose n} \lambda^n\,,
\hspace{0.5cm} Z(N,l)={l-(2m-1)(N+1)\choose N}\,.
\label{omzsim-eq}
\end{equation}
The occupation probabilities $p(l;N,M)$ and $\tilde p(l;\lambda,M)$ in
the canonical and grand-canonical ensemble then follow by inserting
these expressions into eqs.~(\ref{plnm-eq},{\ref{pllmsimp-eq}), and
the correlations analogously. One can show \cite{Buschle:1999} that
$p(l;N,M)$ and $\tilde p(l;\lambda,M)$ become maximal at the points
$l=2m$ and $l=M+1-2m$ closest to the walls. The reason for this is
that by fixing the position of a rod next to a wall the number of
possible configurations (and hence the entropy) for the remaining
$(N-1)$ rods will be largest.

Extending this line of thinking one would guess that the most likely
configuration near a wall is that where the rods are at positions
$l=2m,4m,6m,\ldots$. One then should expect oscillations in the occupation
probabilities to emerge with a period of typical size $2m$. In fact, in the
thermodynamic limit $M\to\infty$ one finds
$\lambda=1/\varphi(s_0)=(1-s_0)s_0^{-2m}$ with $s_0=(1-2m\bar p)/[1-(2m-1)\bar
p]$ (see eq.(\ref{barp-eq})).  Moreover, as shown in Appendix B, the zeros
$s_j=|s_j|\exp(i\theta_j)$ (see Sec.~\ref{specase}) are all different, and
eq.~(\ref{pinf-eq}) becomes
\begin{equation}
p_\infty(l,\lambda)=\bar p + \sum_{j=1}^{2m-1} c_j(\lambda)
\left(\frac{s_0}{|s_j|}\right)^l e^{-i\theta_j l}\,.
\label{pinfhr-eq}
\end{equation}
The $\theta_j$ are in the open interval $0<\theta_j<2\pi$, that means the
profile $p_\infty(l,\lambda)$ is a superposition of simple oscillating and
exponentially decaying functions. When considering a system of finite length,
the effects induced by the second wall at position $M+1-m$ on the profile near
the first wall at position $m$ are of order $l/M$ in the grand-canonical
ensemble and of order $l^2/M$ in the canonical ensemble. This is proven in
Appendix B (more precisely, $l$ must be of order ${\rm o}(M^{1/2})$ in the
canonical and of order ${\rm o}(M)$ in the grand-canonical ensemble to obtain
vanishing contributions in the thermodynamic limit.) Hence, the finite size
corrections to eq.~(\ref{pinfhr-eq}) become small for large $M$.

More surprising, if the number density is smaller than half of that
for the closed packed configuration, i.e. $\bar p<1/4m$, one can show
(see Appendix B) that in the canonical ensemble the $p(l;N,M)$ are
constant ($l$-independent) inside the central region ${\cal
R}_1\equiv\{l_1\in{\mathbb N}\;|\; l^{(1)}\le l_1\le M+1-l^{(1)}\}$
with $l^{(1)}\equiv (2m-1)N+1$.  At the outer boundary points
$l^{(-)}\equiv l^{(1)}-1$ and $l^{(+)}\equiv M+2-l^{(1)}$,
$p(l^\mp;N,M)$ is different from the constant value inside ${\cal
R}_1$, that means ${\cal R}_1$ is ``maximal'' in the sense that there
exits no other region of constant occupation probability enclosing
parts of ${\cal R}_1$. It is interesting to note that
$|p(l^{(\mp)}\pm1;N,M)-p(l^{(\mp)};N,M)|=1/Z(N,M)$, that means the
logarithm of the jump in the occupation probability at the boundaries
of ${\cal R}_1$ provides the free energy $\propto\log
Z(N,M)$. Furthermore, the joint probabilities $p_s(l_1,\ldots,l_s;N,M)$
are translationally invariant inside (``maximal'') regions ${\cal
R}_s\equiv\{ (l_1,\ldots,l_s)\in {\mathbb N}^s)\;|\;l^{(s)}<l_1;2m\le
l_{k}-l_{k-1} \mbox{ for } k=2,\ldots,s; l_s\le M+1-l^{(s)}\}$ with
$l^{(s)}=(2m-1)(N+1-s)+1$,\footnote{By saying that the
$p_s(l_1,\ldots,l_s;N,M)$ are not translationally invariant if
$l_1<\ldots<l_s\notin {\cal R}_s$ we exclude the trivial case, where
$l_1$ and $l_s$ {\it must} be occupied by the first and last rod
center, respectively (i.e. for $l_1<4m$ and $l_s>M+1-4m$). (In this
case, our system can be considered as being composed of $N-2$ rods on
a chain of length $(l_s-l_1)$ and the $p_s(l_1,\ldots,l_s;N,M)$ are
translationally invariant then on trivial reasons.)} that means there
exists a function $f(x_1,\ldots,x_{s-1};N,M)$ such that for all
$(l_1,\ldots,l_s)\in{\cal R}_s$,
$p_s(l_1,\ldots,l_s;N,M)=f(l_2-l_1,\ldots,l_s-l_{s-1};N,M)$.
Corresponding regions have been found in the continuum version of the
hard-rod lattice gas, the so-called hard-core fluid model
\cite{Leff/Coopersmith:1967}. Moreover, if $(l_{j+1}-l_{j})\ge
[(2m-1)(N+1-s)+1]$ for all $j=1,\ldots,s-1$, then
$p_s(l_1,\ldots,l_s;N,M)$ is constant for $(l_1,\ldots,l_s)\in{\cal
R}_s$ (this was shown to hold true in the continuum model for the pair
distribution functions only \cite{Flicker:1968}).  As is shown in
Appendix B also, the situation is quite different in the
grand-canonical ensemble. Here, there exist no regions of constant
occupation probabilities and translational invariance of the joint
probabilities (except for trivial cases$^1$).

\subsection{Free Energy Functional}\label{freeen-subsec}

An alternative way to treat the hard-rod lattice gas has been followed
in \cite{Robledo/Varea:1981}. In this approach, which relates to
density functional theory of classical fluids, one considers the
grand-canonical ensemble and defines the occupation numbers $x_i$,
where $x_i=1$ if site $i$ is occupied by a rod center, and $x_i=0$
else.\footnote{For the following it is convenient to make the
  transformation $l\to l-(2m-1)$ of the site positions and to change
  the system size according to $M\to M+4m-2$.  After these
  replacements the rods at the boundaries are at positions $-2m+1$ and
  $M+2m$ corresponding to hard walls at positions $-m+1$ and $M+m$.
  Accordingly, the possible positions of the rod centers are
  $1,\ldots,M$.} Note that these random variables are not independent:
Since the rods have size $2m$ we have to require $x_j=0$ for
$|j-i|<2m$ if $x_i=1$. We define ${\cal C}_{M}$ as the set of all
allowed configurations $\{x_i\}$.  The idea then is to calculate
explicitly the probability $\chi(x_1,\ldots,x_M)$ of an allowed
configuration $(x_1,\ldots,x_M)\in {\cal C}_{M}$ by regarding the
occupation probabilities $\tilde p_i=\langle x_i\rangle$ as fixed
($\langle[\ldots]\rangle \equiv\sum_{(x_1,\ldots,x_M)\in{\cal C}_{M}}
[\ldots]\chi(x_1,\ldots,x_M)$). For given $\tilde p_i$ it turns out
that $\log\chi(x_1,\ldots,x_M)$ depends linearly on the occupation
numbers (which is a fortunate feature of the hard rod system, see
below).  By equating $\chi(x_1,\ldots,x_M)$ with the Boltzmann formula
for all $(x_1,\ldots,x_M)\in {\cal C}_{M}$ we have
\begin{equation}
-\log\Omega(\lambda,M)=
\log\chi(x_1,\ldots,x_M)+\sum_{s=1}^M x_s [u(s)-\mu]\,,
\label{zfunc-eq}
\end{equation}
where $\mu=\log\lambda$ is the chemical potential.
Taking now the expectation value of (\ref{zfunc-eq}) with respect to the $x_i$,
an exact density functional 
$\beta{\cal F}(\tilde p_1,\ldots,\tilde p_M)=-\log \Omega(\lambda,M)$ of the
$\tilde p_i$ is obtained.

In order to determine $\chi(x_1,\ldots,x_M)$ for
$(x_1,\ldots,x_M)\in{\cal C}_{M}$ we will make use of a Markov
property. (Note that the Markov property is not valid in the canonical
ensemble.)  The constraint given by the finite rod lengths implies
that the conditional probabilities $w_s(x_s|x_{s-1},\ldots x_1)$ for
the occupation number at site $s$ to be $x_s$, given $x_{s-1},\ldots
x_1$, are independent of $x_1,\ldots x_{s-2m}$. In other words
$w_s(x_s|x_{s-1},\ldots x_1)$ fulfills the generalized Markov condition
(which is rather obvious here but can be proven rigorously too
\cite{Buschle:1999})
\begin{equation}
w_s(x_s|x_{s-1},\ldots x_1)=w_s(x_s|x_{s-1},\ldots ,x_{s-2m+1})\,.
\label{markov-eq}
\end{equation}
Due to this property we can express the joint probabilities
$\chi(x_1,\ldots,x_M)$ as
\begin{eqnarray}
\chi(x_1,\ldots,x_M)=&&w_1(x_1)
w_2(x_2|x_1)\ldots w_{2m}(x_{2m}|x_{2m-1},\ldots,x_1)\ldots\nonumber\\
&&\times w_s(x_s|x_{s-1},\ldots,x_{s-2m+1})\ldots 
w_M(x_M|x_{M-1},\ldots,x_{M-2m+1})\,.
\label{markovchain-eq}
\end{eqnarray}
It is convenient to formally extend the system to integers $i\le0$ and
to set $x_i=0$ for all $-2m+2\le i\le0$, such that we can write
eq.~(\ref{markovchain-eq}) in the compact form
\begin{equation}
\chi(x_1,\ldots,x_M)=\prod_{s=1}^M\, w_s(x_s|x_{s-1},\ldots,x_{s-2m+1})\,.
\label{chi1-eq}
\end{equation}

For calculating $w_s(x_s|x_{s-1},\ldots,x_{s-2m+1})$ we have to deal with two
cases only: {\it (i)} One of the given random variables
$x_{s-2m+1},\ldots,x_{s-1}$ is equal to one and the rest of them equal to
zero, and {\it (ii)} all $x_{s-2m+1},\ldots,x_{s-1}$ are zero. In all other
cases there would be at least two of the $x_{s-2m+1},\ldots,x_{s-1}$ equal to
one, but this is not allowed, because it would imply that rods overlap. For
the same reason we must have $x_s=0$ in situation {\it (i)}, that means we
obtain
\begin{equation}
w_s(x_s=0|x_{s-1}=0,\ldots, x_i=1,\ldots,x_{s-2m+1}=0)=1\,.
\label{wsi-eq}
\end{equation}
The situation {\it (ii)} is more complicated. By definition we can write
\begin{equation}
w_s(x_s|0,\ldots,0)=
\frac{\kappa_{s,s-2m+1}(x_s,x_{s-1}=0,\ldots,x_{s-2m+1}=0)}
{\kappa_{s-1,s-2m+1}(x_{s-1}=0,\ldots,x_{s-2m+1}=0)}\,,
\label{wsii1-eq}
\end{equation}
where $\kappa_{l,k}(x_l,\ldots,x_k)$ is the joint probability for the
configuration $\{x_l,\ldots,x_k\}$ to occur. For $k-l<2m$ the normalization
condition yields (again because of the non-overlapping condition)
\begin{equation}
1=\sum_{\{x_j\}} \kappa_{l,k}(x_l,\ldots,x_k)=
\kappa_{l,k}(0,\ldots,0)+\sum_{j=k}^l \kappa_{l,k}(0,\ldots,x_j=1,\ldots,0)\,.
\label{norm-eq}
\end{equation}
By definition we further have (for $k-l<2m$)
\begin{equation}
\tilde p_i=\sum_{\{x_j\}} x_i\, \kappa_{l,k}(x_l,\ldots,x_k)=
\kappa_{l,k}(0,\ldots,x_i=1,\ldots,0)\,,
\label{pi-eq}
\end{equation}
and hence it follows form eqs.~(\ref{norm-eq},\ref{pi-eq}),
\begin{equation}
\kappa_{l,k}(0,\ldots,0)=1-\sum_{j=k}^l \tilde p_j\,.
\label{kappa0-eq}
\end{equation}
Inserting the $\kappa_{l,k}(x_l,\ldots,x_k)$ from
eqs.~(\ref{pi-eq}, \ref{kappa0-eq}) into eq.~(\ref{wsii1-eq}),
we obtain
\begin{equation}
w_s(x_s|0,\ldots,0)=\left\{\begin{array}{l@{\hspace{0.3cm}}l}
\frac{1-t_m(s)}{1-t_m'(s)}\,, & x_s=0\,, \\[0.2cm]
\frac{\tilde p_s}{1-t_m'(s)}\,, & x_s=1\,, \end{array}\right.
\label{wsii2-eq}
\end{equation}
where we have defined $t_m(s)=\sum_{j=0}^{2m-1}\tilde p_{s-j}$ and
$t_m'(s)=\sum_{j=1}^{2m-1}\tilde p_{s-j}=t_m(s)-\tilde p_s$.  The results
(\ref{wsi-eq}, \ref{wsii2-eq}) can be combined to express
$w(x_s|x_{s-1},\ldots,x_{s-2m+1})$ in the general form
\begin{equation}
w(x_s|x_{s-1},\ldots,x_{s-2m+1})=\tilde p_s^{x_s}
\frac{[1-t_m(s)]^{\left(1-\sum_{j=0}^{2m-1} x_{s-j}\right)}}
{[1-t_m'(s)]^{\left(1-\sum_{j=1}^{2m-1} x_{s-j}\right)}}\,.
\label{ws-eq}
\end{equation}

Note that the $x_i$ appear linearly in the exponents of the transition
matrix (\ref{ws-eq}), such that by inserting (\ref{ws-eq}) into
eq.~(\ref{chi1-eq}), and taking the logarithm, we find that $\log\chi$
is {\it linear} in the $x_i$. Using eq.~(\ref{zfunc-eq}) and averaging
over $x_i$ we finally obtain \cite{Robledo/Varea:1981}
\begin{eqnarray}
\beta {\cal F}(\tilde p_1,\ldots,\tilde p_M)&=&
\sum_{s=1}^M \tilde p_s [u(s)-\mu] + \sum_{s=1}^M \tilde p_s \log \tilde p_s
+\\ 
&&{}\sum_{s=1}^M \left(1-t_m(s)\right)\log\left(1-t_m(s)\right) -
\sum_{s=1}^M \left(1-t_m'(s)\right)\log\left(1-t_m'(s)\right)\,.\nonumber
\label{ffunc-eq}
\end{eqnarray}

The functional (\ref{ffunc-eq}) becomes minimal for the equilibrium
density profile $\tilde p_l\equiv \tilde p(l;\lambda,M)$.  The
corresponding system of equations reads ($l=1,\ldots,M$)
\begin{equation}
\frac{\partial(\beta {\cal F})}{\partial \tilde p_l}
(\tilde p_1,\ldots,\tilde p_M)=
-\mu\!+\! u(l)\!+\!\log\tilde p_l\!+\!
\sum_{s=l+1}^{l+2m-1}\log(1-\tilde t_m'(s))\!-\!
\sum_{s=l}^{l+2m-1}\log(1-\tilde t_m(s))=0
\label{struc-eq}
\end{equation}
It is clear that $\tilde p(l;\lambda,M)$ from eq.~(\ref{pllm-eq})
(after making the transformations $l\to l-(2m-1)$ and $M\to M+2-4m$)
must solve (\ref{struc-eq}). From a mathematical point of view this is
an interesting example, where a system of coupled nonlinear difference
equations (eq.~(\ref{struc-eq})) can be mapped by a nonlinear
transformation (eq.~(\ref{pllm-eq})) onto a simple system of
independent linear difference equations (eq.~(\ref{recurom-eq})). For
the special case of a vanishing external potential even an explicit
solution exists (see eqs.~(\ref{pllmsimp-eq}, \ref{omzsim-eq})). A
direct proof that $\tilde p(l;\lambda,M)$ from (\ref{pllm-eq}) indeed
solves (\ref{struc-eq}) is given in Appendix C.

Next we rederive the exact free energy functional of Percus
\cite{Percus:1976} by taking the proper continuum limit of
eq.~(\ref{ffunc-eq}). To do this we first have to note that
eq.~(\ref{ffunc-eq}) gets slightly modified, when it is viewed as
resulting from a discretized form of an originally continuous system.
This continuous system is defined by hard rods of length $\sigma$ with
positions $0<y_i<L$, $y_{i+1}-y_i\ge \sigma$.  In a discretization,
we may subdivide the continuous system into $M$ intervals ${\cal I}_s$
($s=1,\ldots,M$) of equal size $\Delta y=L/M$, and may set the rod
length $2m$ in the new discrete variables $s$ equal to the integer
part of $\sigma L/M$.  The occupation number $x_s$ of the interval
${\cal I}_s$ is defined to be zero, if none of the $y_i\in {\cal I}_s$
and one else. Then the joint probability $q(i_1,\ldots,i_N)$ to find
$N$ rods at positions $y_1,\ldots,y_N$ in the intervals ${\cal
  I}_{i_1},\ldots,{\cal I}_{i_N}$ is,
\begin{eqnarray}
q(i_1,\ldots,i_N)&=&\Omega(\lambda,L)^{-1}
\prod_{k=1}^N\int_{{\cal I}_{i_k}} dy_k
\exp(-[u(y_k)-\mu])\nonumber\\
&=&\Omega(\lambda,L)^{-1}(\Delta y)^N\exp\left(-\sum_{k=1}^N
  [u(i_k)-\mu]
\right)
\left[1+\frac{{\rm o}(\Delta y)}{\Delta y}\right]
\label{qprob-eq}
\end{eqnarray}
Since there is a one-to-one correspondence between the sets
$\{i_1,\ldots,i_N\}$ and $\{x_1,\ldots,x_M\}$ we immediately obtain
\begin{equation}
\chi(x_1,\ldots,x_M)=\Omega(\lambda,L)^{-1}(\Delta y)^{\sum_{s=1}^M x_s}
\exp\left(-\sum_{s=1}^M [u(s)-\mu]\, x_s\right)
\left[1+\frac{{\rm o}(\Delta y)}{\Delta y}\right]
\label{chi2-eq}
\end{equation}
Repeating the steps leading to eq.~(\ref{ffunc-eq}) we get a modified
functional ${\cal F}(\tilde p_1,\ldots,\tilde p_M)$, which is the same as
given in (\ref{ffunc-eq}) plus the term $[-(\sum_s p_s)\log\Delta y+{\rm
  o}(\Delta y)/\Delta y]$.  The $\tilde p_s$ are related to the
occupation number density $\rho(y)$ in the continuous system by
$\tilde p_s=\int_{{\cal I}_s} dy\rho(y)=[\rho(sL/M)\Delta y+{\rm o}(\Delta
y)]$ and by inserting this in the modified form of
eq.~(\ref{ffunc-eq}) we obtain in the limit $M\to\infty$ ($\Delta
y\to0$) the Percus functional
\begin{equation}
\beta {\cal F}[\rho]=\int_0^L dy\, \rho(y) \left\{u(y)-\mu+\log\rho(y)-
\left[1+\log(1-t(y))\right]\right\}\,,
\label{percus-eq}
\end{equation}
where $t(y)=\int_{y-\sigma}^y dz\, \rho(z)$. To consider
eq.~(\ref{percus-eq}) as a mass density functional one should remember
the relation $\rho_{\rm mass}(y)=t(y+\sigma/2)=
\int_{y-\sigma/2}^{y+\sigma/2} dz \rho(z)$ between the mass and the
number density (that might be easily inverted by Laplace
transformation).

\section{Density Profiles and Pair Correlations Near Walls}
\label{numerics-sec}
\setcounter{equation}{0} In this Section we calculate density profiles
and correlations for some cases to exemplify the formalism developed
in the previous Sections~\ref{takahashi-sec} and \ref{hardrod-sec}.

Figure~1 shows the occupation probability $p_\infty(l)$ for a system
of hard rods as a function of the distance $l$ from a hard wall for
{\it (a)} $\bar p=0.1$ and various (half) rod lengths $m=2,3,4$, and
{\it (b)} $\bar p=0.02$ and $m=14,18$, and 22 ($p_\infty(l)$ was
calculated from eqs.~(\ref{recurom-eq},\ref{pinf-eq})).  As can be
seen from the figure, $p_\infty(l)$ exhibits oscillations with a
period of order $2m$, which become more pronounced with increasing
$m$. For large $l$, $p_\infty(l)$ approaches $\bar p$.  Note that for
$\bar p=0.1$ the closed packed situation occurs already at $m=5$ and
the discreteness of the system is important (see Fig.~1a), while $\bar
p=0.02$ (Fig.~1b) corresponds to a continuum situation.  The data in
Fig.~1b indicate that $p_\infty(l)$ in the continuum limit (see
Sect.~\ref{freeen-subsec}) might have a discontinuity in the first
derivative at the first minimum. Indeed this discontinuity occurs and
its origin can be understood from the solution of the discrete system:
From eq.~(\ref{pinf-eq}) and the recursion relation (\ref{recurom-eq})
one derives
\begin{equation}
p_\infty(l)=s_0p_\infty(l-1)+(1-s_0)p_\infty(l-2m)\,.
\label{recurp-eq}
\end{equation}
Accordingly, when $l<4m$, the second term in (\ref{recurp-eq}) is zero up to
the first minimum in $p_\infty(l)$ at $l=4m-1$, and it first contributes when
$l=4m$. The additional contribution from the second term yields the
discontinuity in the first derivative.

The correlation function 
\begin{equation}
C(l)\equiv p_{2,\infty}(2m,l)-p_\infty(2m)p_\infty(l)
\label{cl-eq}
\end{equation}
between the first possible position $2m$ of a rod center and another
rod center that is at distance $l$ from the wall is shown in Fig.~2 for
the same parameters as in Fig.~1. Similar as $p_\infty(l)$, $C(l)$
oscillates as a function of $l$ with a period of order $2m$; the
strength of the oscillations increases with increasing $m$. For large
$l$, the absolute values of $C(l)$ at its local maxima and minima
decrease exponentially with $l$.

Next we calculate density profiles and correlations for more general cases. To
this end we consider {\it (i)} hard rods 
($v(l)=v_{\scriptscriptstyle\rm HR}(l)$)
in the presence of a ``soft wall''
with an attractive potential
\begin{equation}
\beta u_0(l)\equiv -5\exp(-l/20)\,,
\end{equation}
and {\it (ii)} particles with a 
Lennard-Jones type Takahashi interaction of the form
\begin{equation}
\beta v_{\scriptscriptstyle\rm LJ}(l)\equiv \left\{
\begin{array}{r@{\hspace{0.3cm}}l}
\infty & l< 2m\, \\[0.2cm]
-4 & 2m\le l\le 3m\\[0.2cm]
0 & {\rm else}\end{array}\right.
\end{equation}
in the presence of a hard wall ($u(l)=0$). Figure~3 shows $p_\infty(l)$
for these two cases in comparison with the hard rod system for {\it
  (a)} $m=4$ and {\it (b)} $m=18$ (to calculate $p_\infty(l)$ for
$u(l)=u_0(l)$ we have chosen a large system size $M=10^4$ and used
eqs.~(\ref{recurom-eq}, \ref{pllm-eq}).  For both cases oscillations occur
similar as in the hard rod system.  In Fig.~3a the attractive wall
potential causes the maxima and minima to become more pronounced than
in the other cases, while in Fig.~3b only the occupation probability
for the first rod next to the wall is strongly enhanced. Because the
first rod center is strongly attracted by the wall, the position of
the following minima and maxima of $p_\infty$ are shifted toward the
wall.  The weaker effects of the external potential in the
continuum-like situation (Fig.~3b) are due to the fact that the first
minimum of $p_\infty(l)$ occurs at a position, where $u_0(l)$ is
already very small.  For the Lennard-Jones type interaction we find
the oscillations in Fig.~3a to be stronger than in the hard rod
system, but the probability of the first rod to be right at the wall
is reduced (for smaller $m$, however, $p_\infty(2m)$ can be larger
than in the hard rod system).  As can be seen from Figs.~4a,b, the
changes of the correlation functions caused by the external potential
$u_0(l)$ and by the interaction potential $v_{\scriptscriptstyle\rm
  LJ}(l)$ are fully analogous to the changes found for $p_\infty(l)$
in Figs.~3a,b.

\section{Summary}
\label{concl-sec}
\setcounter{equation}{0} 

As demonstrated in Sec.~\ref{numerics-sec}, the recursion relations
derived in Sec.~\ref{takahashi-sec} provide an efficient method to
determine density distributions and correlations in the Takahashi
lattice gas for arbitrary interactions and external potentials. We
have proven in Appendix A that the wall-induced density oscillations
decay exponentially into the bulk, if the interaction potential has
a finite range. If the potential has no finite range but still decays to
zero when the inter-particle distance increases toward infinity, one
can show only that the density for large distances from the wall
converges to a constant bulk value.

In the special case of the hard rod lattice gas, we have rederived the
exact free energy functional of the occupation probabilities and the
associated nonlinear system of coupled nonlinear difference
equations. The general formalism derived in Sec.~\ref{takahashi-sec}
allowed us to give a solution of these nonlinear difference equations
in terms of the solution of a system of independent linear
equations. Furthermore, various central regions between the confining
walls have been specified in the canonical ensemble, where the
occupation probabilities are constant, and where the correlations
functions are translationally invariant or even constant too. In the
grand canonical ensemble it was shown that such regions do not exist
(except for trivial cases, see above).

It is possible to extend the calculations for the hard rod system to
more general situations, as, for example, to systems where certain
groups of particles have differing rod lengths, or even to systems
with randomly distributed rod lengths.

\pagebreak

\begin{center}
{\bf APPENDIX A}\label{appendixa-sec}
\end{center}
\renewcommand{\theequation}{A.\arabic{equation}}
\setcounter{equation}{0}

For $|s|<1$ and $\lambda\in(0,\infty)$ the generating functions of the
canonical and grand-canonical partition functions are (compare with
eq.~(\ref{gls-eq}))
\begin{equation}
H(N,s)=\sum_{M=0}^{\infty} Z(N,M) s^M=\frac{\varphi(s)^{N+1}}{s}\,,\quad
G(\lambda,s)=\sum_{M=0}^{\infty} \Omega(\lambda,M)s^M=
\frac{\varphi(s)}{s[1-\lambda\varphi(s)]}\,,
\end{equation}
with
\begin{equation}
\varphi(s)=\sum_{l=1}^{\infty} \exp[-v(l)] s^l\,,
\label{app-gls-eq}
\end{equation}
where $v(l)$ has the following properties
\begin{equation}
\begin{array}{rll}
{\it (i)} \hspace{0.7cm}& \max\limits_{l_\star\le l<\infty} |v(l)|<\infty & 
\hspace{0.5cm}\mbox{for some}\hspace{0.2cm} l_\star<\infty\,, \\
{\it (ii)} \hspace{0.7cm}& \lim\limits_{l\to\infty} v(l)=0\,. &
\end{array}
\end{equation}

{\bf Lemma A.1.\hspace{0.2cm}} (See also \cite{Lenard:1961}.) For given
$\lambda\in(0,\infty)$, there exists exactly one real positive solution
$s_0(\lambda)\in(0,1)$ of $1-\lambda\varphi(s)=0$ and this root is simple. If
$s_j\in\mathbb{C}$ is another root of $1-\lambda\varphi(s)=0$, then
$|s_j|>s_0$.

{\sl Proof.\hspace{0.2cm}} The series $\sum_{l=1}^{\infty} |\exp[-v(l)] s^l|$
converges for $|s|<1$, and accordingly $\varphi'(s)=\sum_{l=1}^\infty
l\exp[-v(l)] s^{l-1}>0$ for positive real $s$. Since $\varphi(0)=0$ and
$\varphi(1)=\infty$, there exists exactly one $s_0\in(0,1)$, for which
$\varphi(s_0)=1/\lambda$ (with multiplicity one). If $s_j$ were another
root of $1-\lambda\varphi(s)=0$ with $|s_j|<s_0$, then
$|\varphi(s_j)|\le\varphi(|s_j|)<\varphi(s_0)=1/\lambda=\varphi(s_j)$, which
is impossible.  If $s_j=s_0\exp(i\theta)$ with $\theta\in(0,2\pi)$ we have to
require $\sum_{l=1}^{\infty} \exp[-v(l)] s_0^l [1-\cos(l\theta)]=0$ and hence
$\theta=2\pi k$, $k\in\mathbb{Z}$, in contradiction to the assumption
$\theta\in(0,2\pi)$.

\vspace*{0.2cm}
{\bf Theorem A.1.\hspace{0.2cm}} Let $s_0(\lambda)\in(0,1)$ be the unique real
positive solution of $1-\lambda\varphi(s)=0$. Then
\begin{equation}
\Omega(\lambda,M)\sim c_0(\lambda)s_0(\lambda)^{-M}\,,
\hspace{0.6cm}{\rm for}\hspace{0.3cm} M\to\infty\,.
\label{omas-eq}
\end{equation}
where $c_0(\lambda)=\varphi(s_0)^2/s_0^2\varphi'(s_0)$.

{\sl Proof.\hspace{0.2cm}} Let $a_k\equiv\Omega(\lambda,k-1)s_0^k$ and
$f_k\equiv\exp[- v(k)]s_0^k/\sum_{l=1}^{\infty}\exp[-v(l)]s_0^l$, for
$k=1,2,\ldots$, and set $\lambda=1/\varphi(s_0)$. Then, according to
eq.~(\ref{recurom-eq}),
\begin{equation}
a_M=\varphi(s_0) f_M + \sum_{l=1}^{M-1} f_{M-l} a_l=
\sum_{l=1}^M a_{M-l} f_l\,,
\label{am-eq}
\end{equation}
where $a_0\equiv\varphi(s_0)$. Obviously, $f_l\ge0$ and
$\sum_{l=1}^{\infty}f_l=1$. By employing the renewal theorem
\cite{Feller:1968} it follows
\begin{equation}
a_M\sim \frac{a_0}{\sum_{l=1}^\infty lf_l}
\hspace{0.6cm}{\rm for}\hspace{0.3cm} M\to\infty\,.
\label{amas-eq}
\end{equation}
Since $\sum_{l=1}^\infty lf_l=s_0\varphi'(s_0)/\varphi(s_0)$, we obtain
$\Omega(\lambda,M)=$ $a_{M+1} s_0^{-(M+1)}$ $\sim \varphi(s_0)^2 s_0^{-M}
/s_0^2\varphi'(s_0)=c_0(\lambda)s_0^{-M}$ for $M\to\infty$.

\vspace{0.2cm}
{\bf Corollary A.1.\hspace{0.2cm}\vspace{-0.2cm}}
\begin{itemize}
\item[\it (i)] $p_\infty(l;\lambda)\equiv\lim_{M\to\infty} \tilde
  p(l;\lambda,M)=\lambda\Omega(\lambda,l-1)s_0^l$
\item[\it (ii)] $p_{s,\infty}(l_1,\ldots,l_s;\lambda)\equiv
  \lim_{M\to\infty}\tilde p_s(l_1,\ldots,l_s;\lambda,M)=
  p_\infty(l_1;\lambda)\prod_{k=2}^s p_\infty(l_k-l_{k-1};\lambda)$
\item[\it (iii)] $\bar p=\varphi(s_0)/s_0\varphi'(s_0)$, where $\bar
  p=\lim_{M\to\infty} N(\lambda,M)/M$. ($N(\lambda,M)$ is the mean number of
  particles for given fugacity $\lambda$ and $M$.)
\end{itemize}

{\sl Proof.\hspace{0.2cm}} According to eq.~(\ref{pllmsimp-eq}), $\tilde
p(l;\lambda,M)=\lambda\Omega(\lambda,l-1)
\Omega(\lambda,M-l)/\Omega(\lambda,M)$ such that $\tilde p(l;\lambda,M)\sim
\lambda\Omega(\lambda,l-1) c_0(\lambda)s_0^{-(M-l)}/
c_0(\lambda)s_0^{-M}=\lambda\Omega(\lambda,l-1) s_0^l$ by theorem A.1. In
particular, $p_\infty(l;\lambda)=\lambda\Omega(\lambda,l-1)s_0^l$.
Analogously, using eq.~(\ref{psl-eq}),
$p_{s,\infty}(l_1,\ldots,l_s;\lambda)=\lambda^s \Omega(\lambda,l_1-1)
\prod_{k=2}^s \Omega(\lambda,l_k-l_{k-1}-1) s_0^{l_s}$. Since
$l_s=l_1+\sum_{k=2}^s (l_k-l_{k-1})$, we can write
$p_{s,\infty}(l_1,\ldots,l_s;\lambda)=\lambda \Omega(\lambda,l_1-1) s_0^{l_1}
\prod_{k=2}^s [\lambda\Omega(\lambda,l_k-l_{k-1}-1) s_0^{l_k-l_{k-1}}]$, which
together with {\it (i)} gives {\it (ii)}. By definition and theorem A.1,
$N(\lambda,M)=\lambda\partial\log\Omega(\lambda,M)/\partial\lambda \sim
\lambda\partial\log[c_0(\lambda)s_0(\lambda)^{-M}]/\partial\lambda$, from
which follows $\bar p=-\lambda s_0'(\lambda)/s_0(\lambda)$.  But from
$\lambda\varphi(s_0(\lambda))=1$ we immediately obtain $-\lambda
s_0'(\lambda)/s_0(\lambda)=\varphi(s_0)/s_0\varphi'(s_0)$ and hence {\it
  (iii)}.

\vspace*{0.2cm}
{\bf Theorem A.2.\hspace{0.2cm}} Let $N_M$ be any sequence with
$\lim_{M\to\infty} N_M/M=\bar p$. Then
\begin{itemize}
\item[\it (i)] $p_{\infty}(l)\equiv\lim_{M\to\infty}
  p(l;N_M,M)=p_\infty(l;\lambda)= \lambda\Omega(\lambda,l-1)s_0^l$
\item[\it (ii)] $
  \lim_{M\to\infty} p_s(l_1,\ldots,l_s;N_M,M)=
p_{s,\infty}(l_1,\ldots,l_s;\lambda)=
  p_\infty(l_1;\lambda)\prod_{k=2}^s p_\infty(l_k-l_{k-1};\lambda)$\,,
\end{itemize}
where $\lambda=1/\varphi(s_0)$ and $s_0$ is the unique positive solution
of $\varphi(s_0)/s_0\varphi'(s_0)=\bar p$.

{\sl Proof.\hspace{0.2cm}} 
In order to derive the asymptotic limit of the occupation probability
\begin{equation}
p(l;N_M,M)=\sum_{r=1}^l Z(r-1,l-1)\frac{Z(N_M-r,M-l)}{Z(N_M,M)}
\label{plnmappa-eq}
\end{equation}
in the thermodynamic limit, we use (see theorem 6.1 in
\cite{Good:1957})
\begin{equation}
Z(N_M-1,M-1)=\frac{\varphi(s_M)^{N_M}}{\sigma_M s_M^M(2\pi N_M)^{1/2}}
[1+{\rm O}(M^{-1})]\quad {\rm for}\quad  M\to\infty\,,
\label{znmasyma-eq}
\end{equation}
where $\sigma_M^2= \partial_u^2[\log\varphi(s_M e^u)]_{u=0}$ and $s_M$
is the unique non-negative solution of
$N_M/M=\varphi(s)/s\varphi'(s)$.

We further define $\tilde s_M$ as the unique non-negative root of
$(N_M-r)/(M-l)=\varphi(s)/s\varphi'(s)$ (for $r,l$ given
integers) and $\tilde\sigma_M^2= \partial_u^2[\log\varphi(\tilde s_M
e^u)]_{u=0}$. Then it is easy to show that
there exist sequences $\beta_M$ and $\gamma_M$ converging to finite
values for $M\to\infty$ with the property
\begin{equation}
\tilde s_M=s_M\left[1+\frac{\beta_M}{M}+{\rm O}(M^{-2})\right]\,,\quad
\tilde\sigma_M^2=\tilde\sigma_M
           \left[1+\frac{\gamma_M}{M}+{\rm O}(M^{-2})\right]\,.
\label{smsigmam-eq}
\end{equation}
(For example, $\beta_M=(l-r)f(s_M)/s_M f'(s_M)$ with
$f(s)=\varphi(s)/s\varphi'(s)$ has the desired properties.)  Replacing
$N_M$ by $N_M-r$, $M$ by $M-l$, $s_M$ by $\tilde s_M$, as well as
$\sigma_M$ by $\tilde\sigma_M$ in eq.~(\ref{znmasyma-eq}), and using
eq.~(\ref{smsigmam-eq}), we obtain after simple calculations
\begin{equation}
Z(N_M-r-1,M-l-1)=Z(N_M-1,M-1)s_M^l\varphi(s_M)^{-r}[1+{\rm O}(M^{-1})]
\quad{\rm for}\quad M\to\infty\,.
\label{znmasymb-eq}
\end{equation}
Taking the limit $M\to\infty$ in eq.~(\ref{plnmappa-eq}) we thus get
(note that $\lim_{M\to\infty} s_M=s_0$)
\begin{equation}
\lim_{M\to\infty} p(l;N_M,M)=\sum_{r=1}^l Z(r-1,l-1)\varphi(s_0)^{-r}
s_0^l=\lambda\Omega(\lambda,l-1)s_0^l\,.
\end{equation}
The proposition {\it (ii)} follows by using the asymptotic
form of $Z(N_M-r,M-l)$ (eq.~(\ref{znmasymb-eq}) in formula (\ref{ps-eq})).

\vspace*{0.3cm}
\begin{center}
{\bf APPENDIX B}\label{appendixb-sec}
\end{center}
\renewcommand{\theequation}{B.\arabic{equation}}
\setcounter{equation}{0}

For $|s|<1$ and $\lambda\in(0,\infty)$ let
\begin{equation}
G(\lambda,s)=\frac{P(\lambda,s)}{Q(\lambda,s)}=
\sum_{M=0}^{\infty} \Omega(\lambda,M)s^M=\frac{s^{2m-1}}{[1-s-\lambda s^{2m}]},
\label{appb-gls-eq}
\end{equation}
be the generating function from eq.~(\ref{gls-eq}) for the special case of
the hard rod interaction potential defined in eq.~(\ref{hardrodpot-eq}).

\vspace*{0.2cm}
{\bf Lemma B.1.\hspace{0.2cm}} The roots $s_0,\ldots,s_{2m-1}$ of the
polynomial $Q(\lambda,s)$ are all distinct.

{\sl Proof.\hspace{0.2cm}} For one of the roots $s_i$ not to be
simple, we must require that both $Q(\lambda,s_i)=0$ and $(\partial
Q(\lambda,s)/\partial s)_{s=s_i}=0$. But if $(\partial
Q(\lambda,s)/\partial s)_{s=s_i}=0$, we have $s_i^{2m-1}=-1/2m\lambda$ and
inserting this result into $Q(\lambda,s_i)=0$ we obtain
$s_i=2m/(2m-1)>0$, i.e. a positive real number.  On the other hand,
the only real solution of $s_i^{2m-1}=-1/2m\lambda$ is negative, which
is a contradiction. Hence all roots must be simple.

\vspace{0.5cm}
Let
\begin{equation*}
p_\infty(l,\lambda)=\bar p + \sum_{j=1}^{2m-1} c_j(\lambda)
\left(\frac{s_0}{|s_j|}\right)^l e^{-i\theta_j l}
=\sum_{r=1}^{[l/2m]} {l-1-(2m-1)r\choose r-1} (1-s_0)^r
s_0^{l-2mr}
\end{equation*}
be the occupancy probability in the thermodynamic limit, where $[x]$
denotes the integer part of $x$ (see eq.~(\ref{pinfhr-eq}), and
eqs.~(\ref{barp-eq}, \ref{pinf-eq}, \ref{omzsim-eq}), and note that
$\varphi(s_0)=s_0^{2m}/(1-s_0)$).

\vspace*{0.2cm}
{\bf Lemma B.2.\hspace{0.2cm}} For
$\lambda_\infty=1/\varphi(s_0)=(1-s_0)s_0^{-2m}$ with $s_0=(1-2m\bar
p)/[1-(2m-1)\bar p]$, $\bar p=N/M$ ($0<\bar p<1/2m$), 
and $l^2/M^2={\rm o}(M^{-1})$
\begin{itemize}
\item[\it (i)] $p(l;N,M)=p_\infty(l;\lambda_\infty)[1+{\rm O}(l^2/M)]$
\item[\it (ii)] $p_s(l_1,\ldots,l_s=l;N,M)=
p_{s,\infty}(l_1,\ldots,l_s=l;\lambda_\infty)[1+{\rm O}(l^2/M)]$
\end{itemize}

{\sl Proof.\hspace{0.2cm}} 
According to eqs.~(\ref{plnm-eq}),
\begin{equation}
p(l;N,M)=
\sum_{r=1}^N Z(r-1,l-1)\frac{Z(N-r,M-l)}{Z(N,M)}\,.
\label{plnmapp-eq}
\end{equation}
With the definition
\begin{equation}
f(x,y)\equiv \log{\nu M+x\choose\bar p M+y},
\label{fdef-eq}
\end{equation}
where $\nu=1-(2m-1)\bar p$, and $x,y$ are integers,
we can write (see the results for $Z(N,M)$ in eq.~(\ref{omzsim-eq}))
\begin{equation}
\frac{Z(N-r,M-l)}{Z(N,M)}=\exp\left[f\left((2m-1)(r-1)-l,-r\right)-
f\left(-(2m-1),0\right)\right]\,.
\label{zf-eq}
\end{equation}
If $x,y={\rm O}(l)$ we obtain,
by applying Stirling's formula,
$n!=(2\pi n)^{1/2}(n/e)^n\exp[{\rm O}(1/n)]$,
\begin{eqnarray}
f(x,y)&=&
\frac{1}{2}\log\left(\frac{\nu}{2\pi\bar p(\nu-\bar p)M}\right)+
M[\nu\log\nu-(\nu-\bar p)\log(\nu-\bar p)-\bar p\log\bar p]
\nonumber \\
&&{}+x\log\nu-(x-y)\log(\nu-\bar p)-
y\log\bar p+{\rm O}\left(\frac{l^2}{M}\right)
\label{fas-eq}
\end{eqnarray}
Note that the sum over $r$ in eq.~(\ref{plnmapp-eq}) runs at most
up to the integer part of $l/2m$, such that the arguments of the
$f$ functions appearing in eq.~(\ref{zf-eq}) are all of order ${\rm O}(l)$.
Accordingly,
\begin{equation}
\frac{Z(N-r,M-l)}{Z(N,M)}=(1-s_0)^r s_0^{(l-2mr)} [1+{\rm O}(l^{2}/M)]\,,
\label{zf1-eq}
\end{equation}
from which we obtain {\it (i)} by using eq.~(\ref{plnmapp-eq}).

Analogously, starting with eq.~(\ref{ps-eq}),
\begin{equation}
p_s(l_1,\ldots,l_s;N,M)=\hspace{-0.6cm}
\sum_{1\le r_1<\ldots< r_s\le N}\hspace{-0.6cm}
Z(r_1\!-\!1,l_1\!-\!1)
\prod_{k=1}^{s-1} Z(r_{k+1}\!-\!r_k\!-\!1,l_{k+1}\!-\!l_k\!-\!1)
\frac{Z(N\!-\!r_s,M\!-\!l_s)}{Z(N,M)}\,,
\label{psapp-eq}
\end{equation}
and again using eq.~(\ref{zf1-eq}) for the asymptotic behavior, we obtain {\it
  (ii)} after straightforward algebra.

\vspace*{0.2cm}
{\bf Lemma B.3.\hspace{0.2cm}} Let $\lambda$ be the unique fugacity
corresponding to given mean number density $\bar p\in (0,1/2m)$, and
$\lambda_\infty=1/\varphi(s_0)=(1-s_0)s_0^{-2m}$ with $s_0=(1-2m\bar
p)/[1-(2m-1)\bar p]$. Then for $l/M={\rm o}(1)$,
\begin{itemize}
\item[\it (i)] $\tilde p(l;\lambda,M)= p_\infty(l;\lambda_\infty)[1+{\rm
    O}(l/M)]$
\item[\it (ii)] $\tilde p_s(l_1,\ldots,l_s=l;\lambda,M)=
  p_{s,\infty}(l_1,\ldots,l_s=l;\lambda_\infty)[1+{\rm O}(l/M)]$
\end{itemize}

{\sl Proof.\hspace{0.2cm}} According to eq.~(\ref{pllmsimp-eq})
\begin{equation}
\tilde p(l;\lambda,M)=\lambda\Omega(\lambda,l-1)\frac{\Omega(\lambda,M-l)}
{\Omega(\lambda,M)}\,,
\label{pllmsimp-app-eq}
\end{equation} 
where $\Omega(\lambda,M)=\sum_{j=0}^{2m-1}c_j(\lambda)s_j(\lambda)^{-M}$ (see
Lemma B.1 and the discussion after eq.~(\ref{gls-eq})).  In order to find the
asymptotic behavior for $M\to\infty$, one has to keep in mind that $\lambda$
depends on $M$.  Let us denote by $\lambda_M$ the fugacity corresponding to
the mean occupation probability $\bar p$ in a system of (finite) size $M$,
i.e. the unique solution of $\bar p=\lambda
M^{-1}\partial\log\Omega(\lambda,M)/\partial\lambda$ (see Corollary A.1). In
leading order $\lambda_M$ approaches $\lambda_\infty$ as
\begin{equation}
\lambda_M=\lambda_\infty\left[1+\frac{b}{M}+
{\rm O}\left(\frac{1}{M^2}\right)\right]\,,
\label{lambdaM-eq}
\end{equation}
where $b$ is a constant independent of $M$. The proof of this relation can be
worked out by writing $\lambda_M=\lambda_\infty+\epsilon_M$ with
$\lim_{M\to\infty}\epsilon_M=0$,\footnote{Note that, according to Corollary A.1
  we have for $\lambda_\infty=1/\varphi(s_0)$, $\lim_{M\to\infty}
  N(\lambda_\infty,M)/M=\bar p$, i.e. because of the one-to-one correspondence
  between $\bar p$ and $\lambda$ it must hold
  $\lim_{M\to\infty}\lambda_M=\lambda_\infty$.} and inserting this into the
determining equation for $\lambda_M$. Careful expansion of the coefficients
$c_j(\lambda_M)$ and the powers $s_j(\lambda_M)^{-M}$ with respect to
$\epsilon_M$ in the expression for $\Omega(\lambda_M,M)$ then yields
$\lim_{M\to\infty} \epsilon_M M=b\lambda_\infty$ and
$\lim_{M\to\infty}[\epsilon_M-b\lambda_\infty/M]M^2=const.$.

Using eq.~(\ref{lambdaM-eq}) we find
\begin{equation}
\Omega(\lambda_M,M-l)=\exp(b\bar p)\,c_0(\lambda_\infty)\,
s_0(\lambda_\infty)^{-(M-l)}
\left[1+{\rm O}\left(\frac{l}{M}\right)\right]\,,
\label{omegaas-app-eq}
\end{equation}
and thus obtain
$\Omega(\lambda,M-l)/\Omega(\lambda,M)=s_0(\lambda_\infty)^l[1+{\rm
O}(l/M)]$. Since
$\lambda_M\Omega(\lambda_M,l-1)=\lambda_\infty\Omega(\lambda_\infty,l-1)
[1+{\rm O}(l/M)]$, it follows {\it (i)} from
eq.~(\ref{pllmsimp-app-eq}).  Analogously, {\it (ii)} follows by using
eq.~(\ref{psl-eq}) and the asymptotic expansion
(\ref{omegaas-app-eq}).

\vspace{0.2cm} {\bf Theorem B.1.\hspace{0.2cm}} If $(M-1)-(4m-2)N>0$
then the occupation probability $p(l;N,M)$ is independent of $l$ for
all $l\in {\cal R}_1\equiv\left\{l\in{\mathbb N}\;|\;l^{(1)}\le l\le
M+1-l^{(1)}\right\}$, with $l^{(1)}\equiv (2m-1)N+1$, i.e. we can write
$p(l;N,M)=\bar u(N,M)/Z(N,M)$. At the outer boundary points
$l^{(-)}\equiv l^{(1)}-1$ and $l^{(+)}\equiv M+2-l^{(1)}$,
$p(l^{(\mp)};N,M)$ is different from $p(l;N,M)$ inside ${\cal R}_1$,
in particular $p(l^{(\mp)};N,M)=[\bar u(N,M)+(-1)^N]/Z(N,M)$.

{\sl Proof.\hspace{0.2cm}} 
Given $p(l;N,M)$ from eq.~(\ref{plnm-eq}) we first show that
\begin{equation}
u(l;N,M)\equiv\sum_{r=1}^{N}Z(r-1,l-1)Z(N-r,M-l),
\hspace{0.3cm}  Z(r,l)={l-(2m-1)(r+1)\choose r}\,,
\label{ulnm-eq}
\end{equation}
is constant inside ${\cal R}_1$ as long as $(M-1)-(4m-2)N>0$.
To this end we will proof that $u(l;N,M)$ for $l\in{\cal R}_1$
can be rewritten by use of the following combinatorial identity
\begin{eqnarray}
u(l;N,M)&=&\sum_{r=1}^{N}
{l\!-\!1\!-\!(2m\!-\!1)r\choose r\!-\!1}
{M\!-\!l\!-\!(2m\!-\!1)(N\!+\!1\!-\!r)\choose N\!-\!r}\nonumber\\
&=& {M\!+\!1\!-\!4m\!-\!(2m\!-\!1)(N\!-\!1)\choose N\!-\!1}\nonumber\\
&&{}-\sum_{r=2}^{N}(-1)^{r} {2m(r\!-\!1)\!-\!1\choose r\!-\!1}
    {M\!+\!1\!-\!4m\!-\!(2m\!-\!1)(N\!-\!r)\choose N\!-\!r}\nonumber\\
&\equiv& \bar u(N,M)\,,
\label{ucomb-eq}
\end{eqnarray}
which is independent of $l$.

To verify the combinatorial formula we follow the methods described
in the book of Riordan on combinatorial identities \cite{Riordan:1968} and use
the following theorem of Lagrange for implicit functions
\cite{Goulden/Jackson:1983}: Let $\phi(z)$ be a power series in $z$
with $\phi(0)\ne0$, and let $z(t)$ be the unique power series
with $z(0)=0$ satisfying the implicit equation $z(t)=t\phi(z(t))$.
Then, for any power series $F(z)$,
\begin{equation}
\frac{F(z)}{(1-t\Phi'(z))}=
\sum_{n=0}^{\infty}\frac{t^{n}}{n!}
\frac{d^{n}}{d\lambda^{n}}\{F(\lambda)\Phi(\lambda)^{n}\}|_{\lambda=0}\,.
\end{equation}
Applying this theorem to $\phi(z)=(1+z)^{-\beta}$ and $F(z)=(1+z)^\alpha$
with $|z|<1$ and $\alpha,\beta\in\mathbb{N}$ one obtains
\begin{equation}
\frac{(1+z)^{\alpha +1}}{1+(\beta +1)z}=
\sum_{n=0}^{\infty}\omega(\alpha,n)t^{n}\,,
\hspace{0.5cm}
\omega(\alpha,n)=\left\{\begin{array}{c@{\hspace{0.4cm}}l}{\alpha-\beta n
\choose n}\,, & 0\le n\le [\frac{\alpha}{\beta}] \\[0.2cm]
         (-)^{n}{(\beta +1)n-\alpha -1\choose n}\,, & 
n>[\frac{\alpha}{\beta}]\end{array}\right.\,.
\label{lagrange1-eq}
\end{equation}
Using (\ref{lagrange1-eq}) we can write
\begin{equation}
\frac{(1+z)^{\alpha +\gamma +2}}{[1+(\beta +1)z]^2}=
\left[\frac{(1+z)}{1+(\beta +1)z}\right] 
\left[\frac{(1+z)^{\alpha + \gamma+1}}{1+(\beta +1)z}\right]=
\sum_{n=0}^{\infty}\omega(0,n)t^{n} 
\sum_{n=0}^{\infty}\omega(\alpha +\gamma,n)t^{n}\,,
\label{wag1-eq}
\end{equation}
and
\begin{equation}
\frac{(1+z)^{\alpha +\gamma +2}}{[1+(\beta +1)z]^2}=
\left[\frac{(1+z)^{\alpha + 1}}{1+(\beta +1)z}\right]
\left[\frac{(1+z)^{  \gamma+1}}{1+(\beta +1)z}\right]=
\sum_{n=0}^{\infty}\omega(\alpha,n)t^{n} 
\sum_{n=0}^{\infty}\omega(\gamma,n)t^{n}\,.
\label{wag2-eq}
\end{equation}
Comparing (\ref{wag1-eq}) with (\ref{wag2-eq}) and equating expansion
coefficients we obtain
\begin{equation}
\sum_{k=0}^{n}\omega(0,k)\omega(\alpha +\gamma,n-k)=
\sum_{k=0}^{n}\omega(\alpha,k)\omega(\gamma,n-k)\,.
\label{omom1-eq}
\end{equation}
In particular, for $n\le\min\{[\alpha/\beta];[\gamma/\beta]\}$ (see
eq.~(\ref{lagrange1-eq})),\footnote{Equation (\ref{combid1-eq}) may be
  identified as a combinatorial identity derived explicitly in the book of
  Riordan on page~148 \cite{Riordan:1968} by replacing $\beta$ by $-\beta$.
  However, in order to do this one has to define the binomial coefficients for
  negative entries by analytical continuations of the Gamma function.}
\begin{eqnarray}
\sum_{k=0}^{n}{\alpha-\beta k \choose k}{\gamma -\beta (n-k)\choose n-k}&&=
\nonumber \\
&&\hspace{-5cm}{\alpha +\gamma -\beta n\choose n}+ 
\sum_{k=1}^{n}(-1)^{k}{(\beta +1)k-1\choose k}{\alpha +\gamma -\beta
  (n-k)\choose n-k}
\label{combid1-eq}
\end{eqnarray}
By setting $n=N-1$, $k=r-1$, $\beta=2m-1$, $\alpha=l-2m$, and
$\gamma=M+1-2m-l$ for $l\in {\cal R}_1$ (i.e. for $(2m-1)N+1\le l\le
M-(2m-1)N$), we have $N-1\le
\min\{[\frac{l-2m}{2m-1}],[\frac{M+1-2m-l}{2m-1}]\}$, and can use
eq.~(\ref{combid1-eq}) to get eq.~(\ref{ucomb-eq}).

In order to show that ${\cal R}_1$ is a ``maximal'' set, we again use
eq.~(\ref{omom1-eq}) with $\alpha=l^{(\mp)}-2m$, $\beta=2m-1$,
$\gamma=M+1-2m-l^{(\mp)}$, $n=N-1$ and $r=k-1$ (note that
$n-1\le[\alpha/\beta]$, $n>[\alpha/\beta]$, $n\le[\gamma/\beta]$,
$\omega(M+1-2m-l^{(-)},0)=1$ and $\omega(l^{(+)}-2m,N-1)=(-1)^{N-1}$) to
obtain
\begin{equation}
u(l^{(\mp)};N,M)=\bar u(N,M)+(-1)^{N}\,.
\end{equation}
This completes the proof of theorem B.1.

\vspace{0.2cm} {\bf Corollary B.1.\hspace{0.2cm}} For $M+1\ge 6m$ (that means
there can be more than just one rod in the system) there does not exist a
central region where $\tilde p(l;\lambda,M)$ is constant (except for the
trivial set $\tilde{\cal R}_1=\{M/2,M/2+1\}$ for even $M$).

{\sl Proof.\hspace{0.2cm}} Assume that there exist a non-trivial
central region $\tilde{\cal R}_1=\{l_0,\ldots,M+1-l_0\}$ with $2m\le
l_0\le (M-1)/2$ in which $\tilde p(l;\lambda,M)$ is constant. Then
there exists a function
\begin{equation}
f(\lambda,M)=\lambda\Omega(\lambda,l-1)\Omega(\lambda,M-l)
\end{equation}
independent of $l$ for all $l\in\tilde{\cal R}_1$. 

From
\begin{equation}
\lambda\Omega(\lambda,l-1)\Omega(\lambda,M-l)=
\sum_{N=1}^{N_{0}}Z(N,M)p(l,N,M)\lambda^{N}\,,
\end{equation}
for $N_0=\max\{{ N\ge1: Z(N,M)>0}\}=[(M+1)/2m]-1$ (note that
$Z(N,M)={M-(2m-1)(N+1)\choose N}$), we get
\begin{equation}
p(l,N,M)=\frac{1}{Z(N,M)N!}\frac{\partial^{N}\tilde f(\lambda,M)}{\partial^{N}
  \lambda}\Bigl\vert_{\lambda=0}=const. 
\end{equation}
for all $l\in\tilde{\cal R}_1$ and all $1\le N\le N _0$.
But, according to theorem B.1, we have
$p(l,N,M)= const.$ in a central region if and only if
$l\in {\cal R}_1$. Therefore,
\begin{equation}
\tilde{\cal R}_{1}\subseteq {\cal R}_1=
\{(2m-1)N+1,\ldots,M-(2m-1)N\} \hspace{0.2cm}\mbox{for all}\hspace{0.2cm}
1\le N\le N_{0}\,.
\end{equation}
Choosing $N=N_0$ (and because we require $\tilde{\cal R}_1$ to have more than
two elements for excluding trivial situations), we obtain
\begin{equation}
(2m-1)N_0+4\le M+1-(2m-1)N_0\,,
\end{equation}
from which one readily concludes that $M+1\le 6m-1$ in contradiction to the
restriction imposed on $M$.

{\it Comment.} One can even show \cite{Buschle:1999} that there does
not exist {\it any} non-trivial region inside which $\tilde
p(l;\lambda,M)$ is constant.

\vspace{0.2cm} {\bf Theorem B.2.\hspace{0.2cm}} If
$M-1-(4m-2)N+2(m-1)(s-1)>0$ then the joint probabilities
$p_s(l_1,\ldots,l_s;N,M)$ are translationally invariant for
$(l_1,\ldots,l_s)\in {\cal R}_s=\{(l_1,\ldots,l_s)\in{\mathbb N}^s |
l^{(s)}\le l_1; 2m\le l_{k}-l_{k-1} \mbox{
for } k=2,\ldots,s;\, l_{s}\le M+1-l^{(s)}\}$ with
$l^{(s)}\equiv (2m\!-\!1)(N\!+\!1\!-\!s)\!+\!1$,
i.e. there exists a function $f(y_1,\ldots,y_{s-1};N,M)$ exhibiting
the property
\begin{equation}
p_s(l_1,\ldots,l_s;N,M)=f(l_2-l_1,\ldots,l_s-l_{s-1};N,M)\,.
\end{equation}
If $(l_1,\ldots,l_s)\notin {\cal R}_{s}$ and $(l_1,\ldots,l_\tau\pm
  1,\ldots,l_s)\in {\cal R}_s$ for some $\tau\in\{1,\ldots,s\}$ then
\begin{equation}
p_s(l_1,\ldots,l_s;N,M)= f(l_2-l_1,\ldots,l_s-l_{s-1};N,M)+(-1)^{N+1-s}/Z(N,M)\,.
\end{equation}

{\sl Proof.\hspace{0.2cm}} 
The joint probabilities $p(l_{1},\ldots,l_{s};N,M)$ are given by 
(see eq.~(\ref{ps-eq}))
\begin{equation}
p_s(l_1,\ldots,l_s;N,M)=\hspace{-0.7cm}\sum_{1\le r_1<\ldots<r_s\le N}
\hspace{-0.5cm}\frac{Z(r_1\!-\!1,l_1\!-\!1)Z(N\!-\!r_s,M\!-\!l_s)}{Z(N,M)}
\prod_{k=2}^s Z(r_k\!-\!r_{k-1}\!-\!1,l_k\!-\!l_{k-1}\!-\!1)
\end{equation}
By introducing new variables $x_1=r_1$, $x_k=r_k-r_{k-1}$ for
$k=2,\ldots,s$ ($\sum_{k=1}^s x_k=r_s$) we can rewrite this as
\begin{eqnarray}
p_s(l_1,\ldots,l_s;N,M)&=&\hspace{-0.7cm}
\sum_{(x_1,\ldots,x_s)\in{\cal A }_{N,s}}\hspace{-0.5cm}
\frac{Z(x_1\!-\!1,l_1\!-\!1)Z(N\!-\!\sum_{i=1}^s x_i,M\!-\!l_s)}{Z(N,M)}
\prod_{k=2}^s Z(x_k\!-\!1,l_k\!-\!l_{k-1}\!-\!1)\nonumber\\
&=&
Z(N,M)^{-1}\hspace{-1cm}\sum_{(x_2,\ldots,x_s)\in {\cal A }_{N-1,s-1} }
\Bigg\{
\prod_{k=2}^s Z(x_k\!-\!1,l_k\!-\!l_{k-1}\!-\!1)\times\nonumber\\
&&\hspace{1cm}\times\sum_{x_{1}=1}^{N-\sum_{i=2}^{s}x_{i}}
Z(x_1\!-\!1,l_1\!-\!1)Z(N\!-\!\sum_{i=2}^s x_i\!-\!x_1,M\!-\!l_s)\Bigg\}\,,
\label{pcorrcompl-eq}
\end{eqnarray}
where ${\cal A}_{N,s}\equiv\{(x_1,\ldots,x_s)\in{\mathbb N}^s)\;|\;
x_1+\ldots+x_s\le N \}$.

Setting $N'=N-\sum_{i=2}^s x_i$ and $M'=M-(l_s-l_1)$, we have from theorem B.1
(see eq.~(\ref{ulnm-eq}))
\begin{equation}
\sum_{x_1=1}^{N'} Z(x_1\!-\!1,l_1\!-\!1)
Z(N'\!-\!x_1,M'\!-\!l_1)=u(l_1;N',M')\,,
\label{sum-zxl-eq}
\end{equation}
where $u(l_1;N',M')=\bar u(N',M')=const.$ for $(2m-1)N'+1\le l_1\le
M'-(2m-1)N'$. For the $x_i$ this means $(2m-1)(N-\sum_{i=2}^s x_i)+1\le l_1\le
M-(l_s-l_1)-(2m-1)(N-\sum_{i=2}^s x_i)$.  The latter inequality holds true for
all $(x_2,\ldots,x_s)\in{\cal A}_{N-1,s-1}$, if $(2m-1)(N+1-s)+1\le l_1$ and
$l_s\le M-(2m-1)(N+1-s)$, i.e. for $(l_1,\ldots,l_s)\in{\cal R}_s$.
Since $(l_{s}-l_{1})=\sum_{k=2}^{s}(l_{k}-l_{k-1})$ we obtain
\begin{eqnarray}
p(l_{1},\ldots,l_{s};N,M)&=&\hspace{-1.2cm}
\sum_{(x_{2},\ldots,x_{s})\in {\cal A }_{N-1,s-1}
  }\prod_{k=2}^{s}Z(x_{k}\!-\!1,l_{k}\!-\!l_{k-1}\!-\!1)\frac{\bar u(N-\sum_{i=2}^{s}x_{i},M\!-\!\sum_{i=2}^{s}(l_{i}\!-\!l_{i-1}))}{Z(N,M)}\nonumber\\[0.2cm]
& \equiv &  f(l_{2}-l_{1},\ldots,l_{s}-l_{s-1};N,M)\,,
\label{pl1fdef-appb-eq}
\end{eqnarray}
which completes the proof of the first part of the theorem.

To prove the second part, we insert eq.~(\ref{sum-zxl-eq})
in eq.~(\ref{pcorrcompl-eq}) and consider an outer boundary point with
$(l_1,\ldots,l_s)\notin{\cal R}_s$ and
$(l_1,\ldots,l_\tau\pm1,\ldots,l_s)\in{\cal R}_s$ for some
$\tau\in\{1,\ldots,s\}$. In fact, according to the constraints implied by the
finite rod lengths, one can show that only $\tau=1$ and $\tau=s$ are possible.
Due to symmetry we can restrict ourselves to the case $\tau=1$. Then the
outer boundary point is $(l_1=l^{(s)}-1,l_2,\ldots,l_s)$ and we obtain
\begin{equation}
p_s(l_1\!=\!l^{(s)}\!-\!1,\ldots,l_s;N,M)=
\frac{1}{Z(N,M)}\hspace{-0.2cm}\sum_{{\cal A }_{N-1,s-1}}\hspace{-0.2cm}
\left\{\prod_{k=2}^s Z(x_k\!-\!1,l_k\!-\!l_{k-1}\!-\!1)\,u(l_1;N',M')
\right\}_{l_1=l^{(s)}-1}
\label{pl1op-appc-eq}
\end{equation}
Except for the particular configuration $(x_2=1,\ldots,x_s=1)$, all
configurations $(x_2,\ldots,x_s)\in{\cal A }_{N-1,s-1}$ yield
arguments $(l_1,N',M')$ for which $u(l_1,N',M')$ is constant (see the
discussion above). For $(x_2=1,\ldots,x_s=1)$, $l_1$ is an outer
boundary point of the set ${\cal R}_1$ corresponding to a system of size
$M'$ with $N'$ rods. According to theorem B.1 we thus have
$u(l_1,N',M')=\bar u(N',M')+(-1)^{N'}$. By inserting these results in
eq.~(\ref{pl1op-appc-eq}) and by using the definition of
$f(y_1,\ldots,y_{s-1};N,M)$ in eq.~(\ref{pl1fdef-appb-eq}) we obtain
\begin{equation}
p_s(l_1\!=\!l^{(s)}\!-\!1,\ldots,l_s;N,M)=
f(l_2-l^{(s)}\!+\!1,\ldots,l_s-l_{s-1};N,M)+\frac{ (-1)^{N+1-s}}{Z(N,M)}\,.
\end{equation}

\vspace{0.2cm} {\bf Corollary B.2.\hspace{0.2cm}} For $(l_1,\ldots,l_s)\notin
{\cal C}_s\equiv\{(l_1,\ldots,l_s)\in{\mathbb N}^s\;|\; 2m\le l_1< 4m,
2m\le l_k\!-\!l_{k-1} \mbox{ for } k=1,\ldots,s; M\!+\!1\!-\!4m< l_s\le
M\!+\!1\!-\!2m \}$ (see footnote 1 in Sect.~\ref{freeen-subsec}) there does
not exist a region, where $\tilde p_s(l_1,\ldots,l_s;\lambda,M)$ is
translationally invariant.

{\sl Proof.\hspace{0.2cm}} 
If
$\tilde p_s(l_1,\ldots,l_s;\lambda,M)=\tilde f(l_2-l_1,\ldots,l_s-l_{s-1};
\lambda,M)$ for some $(l_1,\ldots,l_s)\notin{\cal C}_s$ then, from 
eq.~(\ref{psl-eq}),
\begin{equation}
\Omega(\lambda,l_1-1)\Omega(\lambda,M'-l_1)=
\frac{\Omega(\lambda,M)}{\lambda^s}
\tilde f(l_2-l_1,\ldots,l_s-l_{s-1};\lambda,M)
\prod_{k=2}^s\frac{1}{\Omega(\lambda,l_k-l_{k-1}-1)}\,,
\end{equation}
where $M'=M-(l_s-l_1)$. Accordingly, there exists a range of consecutive $l_1$
values, where $\Omega(\lambda,l_1-1)\Omega(\lambda,M'-l_1)$ is independent of
$l_1$. This, however, is impossible due to the comment after Corollary B.1.

\vspace{0.2cm} {\bf Theorem B.3.\hspace{0.2cm}} If
$M-2(2m-1)(N+1-s)\ge0$ then the joint probabilities
$p_s(l_1,\ldots,l_s;N,M)$ are constant functions for
$(l_1,\ldots,l_s)\in{\cal B}_s\equiv\{(l_1,\ldots,l_s)\in{\mathbb
N}^s\;|\; (2m-1)(N+1-s)+1\le l_1\,;\,(2m-1)(N+1-s)+1\le l_k-l_{k-1}
\mbox{ for } k=2,\ldots,s\,;\,l_s\le M-(2m-1)(N+1-s)\}$, i.e. we can
write $p_s(l_1,\ldots,l_s;N,M)=\bar v(N,M)$. If
$(l_1,\ldots,l_s)\notin{\cal B}_s$ and
$(l_1,\ldots,l_\tau\pm1,\ldots,l_s)\in{\cal B}_s$ for some
$\tau\in\{1,\ldots,s\}$ then
\begin{equation}
p_s(l_1,\ldots,l_s;N,M)=\bar v(N,M) + \frac{(-1)^{N+1-s}}{Z(N,M)}\,.
\label{pl1lsnm-eq}
\end{equation}

{\sl Proof.\hspace{0.2cm}} 
The explicit formula
(see eqs.~(\ref{ps-eq},\ref{omzsim-eq})
\begin{eqnarray}
Z(N,M)\,p_s(l_1,\ldots,l_s;N,M)&=&\hspace{-0.7cm}
\sum_{(x_1,\ldots,x_s)\in{\cal
    A}_{N,s}}\hspace{-0.3cm}
{l_1\!-\!1\!-\!(2m\!-\!1)x_1\choose x_1\!-\!1}
\times \\
&&\hspace{-3.5cm}
{M\!+\!1\!-\!2m\!-\!l_{s}\!-\!(2m\!-\!1)(N-\!\sum_{i=1}^{s}x_{i})\choose 
      N\!-\!s\!-\!\sum_{i=1}^{s}x_{i}}
\prod_{k=2}^{s}
{l_{k}\!-\!l_{k-1}\!-\!1\!-\!(2m\!-\!1)( x_{k}\!-\!x_{k-1})\choose
        x_{k}\!-\!x_{k-1}\!-\!1}\nonumber\,.
\label{pl1lsnmapp-eq}
\end{eqnarray}
can be rewritten by using eq.~(\ref{lagrange1-eq}) derived in theorem B.1
(remember that we defined ${\cal A}_{N,s}=\{(x_1,\ldots,x_s)\in{\mathbb
  N}^s)\;|\; x_1+\ldots+x_s\le N \}$ after eq.~(\ref{pcorrcompl-eq})). We have
for $\alpha_i\ge 0$, $i=1,\ldots,s+1$, and ${\cal
  E}_{n,s}\equiv\{(n_1,\ldots,n_s)\in{\mathbb N}_0^s\;|\; n_1+\ldots+n_s=n\}$,
\begin{equation}
\frac{(1+w)^{\alpha_{1}+\ldots +\alpha_{s+1}+s+1}}{[1+(\beta
  +1)w]^{s+1}}=\prod_{k=1}^{s+1}
\sum_{n_{k}=0}^{\infty}\omega(\alpha_{k},n_{k})t^{n_{k}}=
\sum_{n=0}^{\infty}t^{n}\sum_{(n_{1},\ldots,n_{s+1})\in{\cal E}_{n,s+1}}
        \prod_{k=1}^{s+1}\omega(\alpha_{k},n_{k})\,
\label{wplus1-eq}
\end{equation}
and
\begin{eqnarray}
\frac{(1+w)^{\alpha_{1}+\ldots +\alpha_{s+1}+1}}{[1+(\beta
  +1)w]^{1}}&&\hspace*{-1cm}\frac{(1+w)^{s}}{[1+(\beta
  +1)w]^{s}}
\!=\!\sum_{n_{1}=0}^{\infty}\omega(\alpha_{1}+\ldots+\alpha_{s+1},n_{1})t^{n_{1}}\prod_{k=2}^{s+1}
\sum_{n_{k}=0}^{\infty}\omega(0,n_{k})t^{n_{k}}\nonumber\\
&\!=\!&\sum_{n=0}^{\infty}t^{n}
\sum_{{\cal E}_{n,s+1}}\hspace{-0.2cm}
\omega(\alpha_{1}+\ldots+\alpha_{s+1},n_{1})\prod_{k=2}^{s+1}\omega(0,n_{k})\,.
\label{wplus2-eq}
\end{eqnarray}
Comparing (\ref{wplus1-eq}) and (\ref{wplus2-eq}) and equating coefficients
we obtain
\begin{equation}
\sum_{(n_{1},\ldots,n_{s+1})\in{\cal E}_{n,s+1}}
\prod_{k=1}^{s+1}\omega(\alpha_{k},n_{k})=\hspace{-0.6cm}
\sum_{(n_{1},\ldots,n_{s+1})\in{\cal E}_{n,s+1}}\hspace{-0.8cm}
\omega(\alpha_{1}+\ldots+\alpha_{s+1},n_{1})\,
\prod_{k=2}^{s+1}\omega(0,n_{k})\,,
\label{sumprod-eq}
\end{equation}
which, for $n\le\min_{1\le k\le s+1}\{[\alpha_k/\beta]\}$ (see
eq.~(\ref{lagrange1-eq})), yields
\begin{equation}
\sum_{{\cal E}_{n,s+1}} \prod_{k=1}^{s+1} {\alpha_k\!-\!\beta n_k\choose n_k} 
= \sum_{{\cal E}_{n,s+1}}
{\alpha_{1}\!+\!\ldots\!+\!\alpha_{s+1}\!-\!\beta n_{1}\choose
    n_{1}}\prod_{k=2}^{s+1}(-1)^{n_{k}}{ (\beta \!+\!1)n_{k}\!-\!1\choose
    n_{k}}\,.
\label{binom-appb-eq}
\end{equation}
We have defined ${-1\choose0}=1$ on the right side here (but not on the
left hand side).

By choosing $\alpha_1=l_1-2m$, $\alpha_k=l_k-l_{k-1}-2m$ for $k=2,\ldots,s$,
$\alpha_{s+1}=M+1-l_s-2m$, $\beta=2m-1$ and $n=N-s\le\min_{1\le k\le
  s+1}\{\alpha_k/\beta\}$, we can identify the combinatorial expression
(\ref{pl1lsnmapp-eq}) by the left hand side of (\ref{binom-appb-eq}).
Substituting the right hand side we then obtain 
\begin{eqnarray}
p_s(l_1,\ldots,l_s;N,M)
&=&Z(N,M)^{-1}\hspace{-0.6cm}
\sum_{(x_1,\ldots,x_s)\in{\cal A}_{N,s}}\hspace{-0.3cm}
(-1)^{N\!-\!\sum_{i=1}^{s}x_{i}}\
{2m(N\!-\!\sum_{i=1}^{s}x_{i})\!-\!1\choose N\!-\!\sum_{i=1}^{s}x_{i}}
  \times \nonumber\\
&&\hspace{3cm}{M\!-\!2ms\!-\!(2m\!-\!1)x_{1}\choose
              x_{1}\!-\!1}\prod_{k=2}^{s}(-1)^{x_{k}\!-\!1}
    {2m(x_{k}\!-\!1)\!-\!1\choose x_{k}\!-\!1}\nonumber\\[0.4cm]
&\equiv& \bar v(N,M)\,,
\end{eqnarray}
which completes the proof of the first part of the theorem.

Now, let $(l_1,\ldots,l_s)\notin {\cal B}_s$ and
$(l_1,\ldots,l_\tau\pm1,\ldots,l_s)\in {\cal B}_s$ for a given
$\tau\in\{1,\ldots,s\}$, and $l_0=0$ and $l_{s+1}=M+1$.  Then $N-s\le
[\alpha_k/(2m-1)]$ for $k\ne\tau$ and $N-s-1=[\alpha_\tau/(2m-1)]$,
which can be shown by simple topological considerations. Accordingly,
for $(n_1,\ldots,n_{s+1})\in{\cal E}_{N-s,s+1}'\equiv{\cal
E}_{N-s,s+1}\backslash (n_1=0,\ldots,n_\tau=N-s,\ldots,n_{s+1}=0)$,
$\omega(\alpha_k,n_k)={\alpha_k-(2m-1)n_k\choose n_k}$, while for
the particular element $(n_1=0,\ldots,n_\tau=N-s,\ldots,n_{s+1}=0)$,
$\omega(\alpha_k,n_k)=(1-\delta_{k\tau})+(-1)^{(N-s)}\delta_{k\tau}$.
Evaluating the left hand side of eq.~\ref{sumprod-eq} we arrive at
\begin{eqnarray}
\sum_{(n_{1},\ldots,n_{s+1})\in{\cal E}_{N-s,s+1}}
            \prod_{k=1}^{s+1}\omega(\alpha_{k},n_{k})&=&
\sum_{(n_{1},\ldots,n_{s+1})\in{\cal E}_{N-s,s+1}'}
 \prod_{k=1}^{s+1}{\alpha_{k}-(2m-1)n_{k}\choose n_{k}} 
                           +(-1)^{N-s}\nonumber\\[0.2cm]
&=& Z(N,M)p_s(l_1,\ldots,l_s;N,M)+(-1)^{N-s}\,.
\label{sumcal-eq}
\end{eqnarray}
On the other hand, the right hand side of eq.~(\ref{sumprod-eq}) is
equal to $Z(N,M)\bar v(N,M)$ and hence we obtain
\begin{equation}
p_s(l_1,\ldots,l_s;N,M)=\bar v(N,M)+\frac{(-1)^{N+1-s}}{Z(N,M)}\,.
\end{equation}

\vspace*{0.3cm}
\begin{center}
{\bf APPENDIX C}\label{appendixc-sec}
\end{center}
\renewcommand{\theequation}{C.\arabic{equation}}
\setcounter{equation}{0}

For $0\le M'+\alpha\le M$ and $\alpha\ge 0$, $\Omega(\lambda;M',\alpha)$ 
is defined as follows (for $v(n)=v_{\scriptscriptstyle\rm HR}(n)$,
see Sect.~\ref{hardrod-sec})
\begin{eqnarray}
\Omega(\lambda;M',\alpha)&=&0,\mbox{ for }  M'< 2m-1 \nonumber\\
\Omega(\lambda;M',\alpha)&=&1,\mbox{ for } 2m-1\le M'< 4m-1 \\
\Omega(\lambda;M',\alpha)&=&1 +\lambda
\sum_{r=2m}^{M'+1-2m}e^{-u(\alpha+r)}\Omega(\lambda;r-1,\alpha), 
\mbox{for} M'\ge 4m-1\nonumber
\end{eqnarray}
\vspace{0.2cm} {\bf Theorem C.1.\hspace{0.2cm}} 
The occupation probability
\begin{equation}
\tilde p(l;\lambda,M)=\lambda e^{-u(l)}
\frac{\Omega(\lambda;l-1,0)\Omega(\lambda;M-l,l)}{\Omega(\lambda;M,0)}
\label{pllM-appc-eq}
\end{equation}
is the unique solution of the following set of nonlinear coupled
difference equations
\begin{equation}
0=
-\log{\lambda}+u(l)+\log\tilde p_l\!
+\!\sum_{s=l+1}^{l+2m-1}\log[1\!-\!\tilde t_m(s)\!+\!\tilde
p_s]
\!-\!\sum_{s=l}^{l+2m-1}\log[1\!-\!\tilde t_m(s)],
\label{diff-appc-eq}
\end{equation}
for $l\in\{2m,\ldots,M\!+\!1\!-\!2m\}$ and $\tilde p_l=0$ else,
where $\tilde t_m(s)=\sum_{j=0}^{2m-1}\tilde p_{s-j}$.

\vspace*{0.2cm}
{\bf Lemma C.1.\hspace{0.2cm}}
Let $M'\ge 2m$. If $0\le M'+\alpha\le M$ and $\alpha\ge0$, then
$\Omega(\lambda;M',\alpha)$ obeys the recursion relations
\begin{itemize}
\item[\it (i)] $\Omega(\lambda;M',\alpha)=
\Omega(\lambda;M'-1,\alpha)+\lambda^{-u(\alpha+M'-2m+1)}
\Omega(\lambda;M'-2m,\alpha)$,
\item[\it (ii)] $\Omega(\lambda;M',\alpha)=
\Omega(\lambda;M'-1,\alpha+1)+\lambda e^{-u(\alpha+2m)}
\Omega(\lambda;M'-2m,\alpha+2m)$.
\end{itemize}

\vspace*{0.2cm}
{\bf Lemma C.2.\hspace{0.2cm}} 
For $s\ge 2m $ let 
\begin{equation}
1-\tilde t_{m}(s)=
\frac{\Omega(\lambda;s-1,0)}{\Omega(\lambda;M,0)}\phi(\lambda;M,s)\,.
\end{equation}
Then $\phi(\lambda;M,s)$ obeys for
$\tilde p_s\equiv\tilde p(s;\lambda,M)$ (from eq.~(\ref{pllM-appc-eq}))
the relations
\begin{itemize}
\item[\it (i)] $\phi(\lambda;M,s)=\Omega(\lambda;M-s+2m-1,s-2m+1)$,
\item[\it (ii)] $\phi(\lambda;M,s-1)=\phi(\lambda;M,s)+ \lambda
e^{-u(s)}\Omega(\lambda;M-s,s)$.
\end{itemize}

\vspace*{0.2cm}
{\sl Proof of Lemma C.1.\hspace{0.2cm}}
For $M'\ge 4m-1$
\begin{eqnarray}
\Omega(\lambda;M',\alpha)&=&1 +\lambda
\sum_{r=2m}^{M'-2m}e^{-u(\alpha+r)}\Omega(\lambda;r-1,\alpha)+ \lambda
e^{-u(\alpha+M'-2m+1)}\Omega(\lambda;M'-2m,\alpha)\nonumber\\
&=& \Omega(\lambda;M'-1,\alpha)+
e^{-u(\alpha+M'-2m+1)}\Omega(\lambda;M'-2m,\alpha)\,.
\label{olma-appc-eq}
\end{eqnarray}
If $2m\le M'<4m-1$, then $\Omega(\lambda;M'-1,\alpha)=1$ and
$\Omega(\lambda;M'-2m,\alpha)=0$. Hence, eq.~(\ref{olma-appc-eq})
is also valid for $2m\le M'<4m-1$, which implies {\it (i)}.

According to (\ref{z-eq}),
\begin{eqnarray}
Z(N,M',\alpha)
&=&\sum_{1\le i_{1},\ldots<i_{N}\le M'}\hspace{-0.8cm}
\exp- \left[\sum_{k=1}^{N}u(\alpha\!+\!i_{k})\!+
\!v(i_{1})\!+\!\sum_{k=2}^{N}v(i_{k}\!-\!i_{k-1})\!+
\!v(M'\!+\!1\!-\!i_{N})\right]\nonumber\\
&=&\sum_{l=1}^{M'} e^{-u(\alpha+l)\!-\!v(l)}\times\nonumber\\
 &&\hspace{0.3cm} \sum_{l\!+\!1\le i_{2},\ldots<i_{N}\le
    M'}\hspace{-0.8cm}
  \exp-\left[\sum_{k=2}^{N}u(\alpha\!+\!i_{k})\!+\!v(i_{2}\!-\!l)\!
   +\!\sum_{k=3}^{N}v(i_{k}\!-\!i_{k-1})\!+
    \!v(M'\!+\!1\!-\!i_{N})\right]\nonumber\\
&=&\sum_{l=1}^{M'}\exp-\left[u(\alpha \!+\!l)\!+\!v(l)\right]\times\nonumber\\
 &&\hspace{-0.2cm} \sum_{1\le j_{1}\!<\!\ldots \!<\!j_{N-1}\le M'\!-\!l}
\hspace{-0.8cm} \exp\!-\!\left[\sum_{k=1}^{N-1}u(\alpha \!+\!l\!+\!j_{k})\!
+\!v(j_{1})\!+\!\sum_{k=2}^{N-1}v(j_{k}\!-\!j_{k-1})\!
+\!v(M'\!-\!l\!+\!1\!-\!i_{N})\right]\nonumber\\
&=&\sum_{l=1}^{M'}e^{-u(\alpha +l)}e^{-v(l)}Z(N-1,M'-l,\alpha+l)\,.
\end{eqnarray}
Remembering that $Z(0,M',\alpha)=e^{-v(M'+1)}$
(see the remark right after eq.~(\ref{recurz-eq})), it holds
\begin{eqnarray}
\Omega(\lambda,M',\alpha)&=&
\sum_{N=0}^{\infty}Z(N,M',\alpha)\lambda^{N}\nonumber\\
&=&e^{-v(M'+1)}+\lambda\sum_{l=1}^{M'}
e^{-u(\alpha +l)}e^{-v(l)}\Omega (\lambda,M'-l,\alpha+l)\,,
\end{eqnarray}
where $v(n)=\infty$ for $n<2m$, and $v(n)=0$ for $n\ge 2m$.
Since $\Omega(\lambda,M'-l,\alpha+l)=0$ for $M'-l<2m-1$ we have
\begin{equation}
\Omega (\lambda,M',\alpha)=1+\lambda\sum_{l=2m}^{M'}e^{-u(\alpha +l)}\Omega
(\lambda,M'-l,\alpha+l)\,,
\end{equation}
which for $r=M'-l$ yields
\begin{equation}
=1+\lambda \sum_{r=2m-1}^{M'-2m}
e^{-u(\alpha+M'-r)}\Omega(\lambda;r,\alpha+M'-r)\,.
\end{equation}
Hence we finally obtain {\it (ii)},
\begin{eqnarray}
\Omega (\lambda,M',\alpha)&=&1+\lambda \sum_{r=2m\!-\!1}^{M'\!-\!2m}
e^{-u(\alpha+M'-r)}\Omega(\lambda;r,\alpha\!+\!M'\!-\!r)\nonumber\\
&=&1+\lambda \sum_{r=2m\!-\!1}^{M'\!-\!1\!-\!2m}
e^{-u((\alpha+1)\!+\!(M'\!-\!1)\!-\!r)}
\Omega(\lambda;r,(\alpha+1)\!+\!(M'\!-\!1)\!-\!r)\nonumber\\
&&\hspace{0.4cm}+\lambda
e^{-u(\alpha+2m)}\Omega(\lambda;M'\!-\!2m,\alpha+2m)\nonumber\\
&=&\Omega (\lambda,M'\!-\!1,\alpha+1)+\lambda 
e^{-u(\alpha+2m)}\Omega(\lambda;M'\!-\!2m,\alpha+2m)\,.
\end{eqnarray}

\vspace*{0.2cm}
{\sl Proof of Lemma C.2.\hspace{0.2cm}} We prove the
proposition {\it (i)} by complete induction with respect to $s$.  For $s=2m$
we have
\begin{equation}
\phi(\lambda;M,2m)=
\frac{\Omega(\lambda;M,0)}{\Omega(\lambda;2m-1,0)}[1-\tilde  t_m(2m)]\,.
\end{equation}
With $ \Omega(\lambda;2m-1,0)=1$, $\tilde t_m(2m)=\sum_{j=0}^{2m-1}\tilde
p_{2m-j}=\tilde p_{2m}$ (note that $\tilde p_{i}=0$ for $i<2m$, and $\tilde
p_{2m}=\lambda e^{-u(2m)}\Omega(\lambda;M-2m,2m)/\Omega(\lambda;M,0)$)
it follows
\begin{equation}
\phi(\lambda;M,2m)=\Omega(\lambda;M,0)-\lambda 
e^{-u(2m)}\Omega(\lambda;M-2m,2m)\,.
\end{equation}
For $M'=M$ and $\alpha=0$ we then obtain by using Lemma C.1. {\it (ii)}
\begin{equation}
\Omega (\lambda,M,0)=\Omega (\lambda,M-1,1)+
\lambda e^{-u(2m)}\Omega(\lambda;M-2m,2m)
\end{equation}
and hence
\begin{equation}
\phi(\lambda;M,2m)=\Omega (\lambda,M,0)-
\lambda e^{-u(2m)}\Omega(\lambda;M-2m,2m)=\Omega (\lambda,M-1,1)\,.
\end{equation}
Accordingly, proposition {\it (i)} is valid for $s=2m$.
Let us now assume that {\it (i)} holds true for $s-1\ge 2m-1$.

Since $\tilde t_{m}(s)=\sum_{j=0}^{2m-1}\tilde p_{s-j}$ and $\tilde
p_{l}=\lambda e^{-u(l)}\Omega(\lambda;l-1,0)\Omega(\lambda;M-l,l)/
\Omega(\lambda;M,0)$ we can write
\begin{eqnarray}
\phi(\lambda;M,s)
&=&\frac{\Omega(\lambda;M,0)}{\Omega(\lambda;s\!-\!1,0)}
\left[1\!-\!\sum_{j=0}^{2m\!-\!1}e^{-u(s\!-\!j)}\lambda 
\frac{\Omega(\lambda;s\!-\!j\!-\!1,0)\Omega(\lambda;M\!-\!s\!+\!j,s\!-\!j)}
     {\Omega(\lambda;M,0)}\right]\nonumber\\
&=&\frac{1}{\Omega(\lambda;s\!-\!1,0)}
\left[\Omega(\lambda;M,0)\!-\!\lambda \sum_{j=0}^{2m\!-\!1}e^{-u(s\!-\!j)}
\Omega(\lambda;s\!-\!j\!-\!1,0)
\Omega(\lambda;M\!-\!s\!+\!j,s\!-\!j)\right]\nonumber\\
&=&\frac{1}{\Omega(\lambda;s\!-\!1,0)}
\left[\Omega(\lambda;M,0)\!-\!\lambda
 \sum_{j=0}^{2m\!-\!1}e^{-u(s\!-\!1\!-\!j)}
\Omega(\lambda;s\!-\!j\!-\!2,0)
\Omega(\lambda;M\!-\!s\!+\!1\!+\!j,s\!-\!1\!-\!j)\right]\nonumber\\
&&{}+\frac{\lambda}{\Omega(\lambda;s\!-\!1,0)}\left[e^{-u(s\!-\!2m)}
\Omega(\lambda;s\!-\!2m\!-\!1,0)
\Omega(\lambda;M\!-\!s\!+\!2m,s\!-\!2m)\right.\nonumber\\
&&\hspace{3.1cm}\left.-e^{-u(s)}
\Omega(\lambda;s\!-\!1,0)\Omega(\lambda;M\!-\!s,s)\right]\nonumber\\
&=&\frac{1}{\Omega(\lambda;s\!-\!1,0)}
\Bigl[\Omega(\lambda;s\!-\!2,0)\phi(\lambda;M,s\!-\!1)\\
&&{}+\lambda
  e^{-u(s\!-\!2m)}\Omega(\lambda;s\!-\!2m\!-\!1,0)
\Omega(\lambda;M\!-\!s\!+\!2m,s\!-\!2m)\Bigr]\!-\!
\lambda e^{-u(s)}\Omega(\lambda;M\!-\!s,s)\nonumber
\end{eqnarray}
where the line 3 follows from line 2 after some straightforward
manipulation of the sum over $j$, and line 4 from line 3 by definition
of $\phi(\lambda;M,s-1)$.

Due to the induction hypothesis, $\phi(\lambda;M,s\!-\!1)=
\Omega(\lambda;M\!-\!s\!+\!2m,s\!-\!2m)$, from which follows
\begin{eqnarray}
\phi(\lambda;M,s)
&=&\frac{\Omega(\lambda;M\!-\!s\!+\!2m,s\!-\!2m)}
{\Omega(\lambda;s\!-\!1,0)}
\left[\Omega(\lambda;s\!-\!2,0)+
\lambda e^{-u(s\!-\!2m)}\Omega(\lambda;s\!-\!2\!m\!-\!1,0)\right]\nonumber\\
&&\hspace{0.5cm}-\lambda e^{-u(s)}\Omega(\lambda;M\!-\!s,s)\,.
\end{eqnarray}
By using {\it (i)} of Lemma C.1. for $M'=s-1$ and $\alpha=0$
we obtain
\begin{equation} 
\Omega(\lambda;s-1,0)=
\Omega(\lambda;s-2,0)+\lambda e^{-u(s-2m)}\Omega(\lambda;s-2m-1,0)
\end{equation}
and hence
\begin{equation}
\phi(\lambda;M,s)=
\Omega(\lambda;M-s+2m,s-2m)-\lambda e^{-u(s)}\Omega(\lambda;M-s,s)
\end{equation}
By using {\it (ii)} of Lemma C.1. for $M'=M-s+2m$ and $\alpha=s-2m$ we find
\begin{equation}
\Omega(\lambda;M-s+2m,s-2m)=\Omega(\lambda;M-s+2m-1,s-2m+1)+
\lambda e^{-u(s)}\Omega(\lambda;M-s,s)
\end{equation}
and hence
\begin{eqnarray}
\Omega(\lambda;M-s+2m,s-2m)-\lambda e^{-u(s)}\Omega(\lambda;M-s,s)
&=&\Omega(\lambda;M-s+2m-1,s-2m+1)\nonumber\\
&=&\phi(\lambda;M,s)\,.
\end{eqnarray}
This completes the proof of part {\it (i)} of Lemma C.2.

To proof the second part {\it (ii)}, we apply {\it (ii)} of Lemma C.1. For
$M'=M-s+2m$ and $\alpha=s-2m$ we obtain
\begin{equation}
\Omega(\lambda;M-s+2m,s-2m)=
\Omega(\lambda;M-s+2m+1,s-2m+1)+\lambda e^{-u(s)}\Omega(\lambda;M-s,s)\,.
\label{omla-appc-eq}
\end{equation}
Due to part {\it (i)} of the Lemma just proven,
$\phi(\lambda;M,s)=\Omega(\lambda;M-s+2m-1,s-2m+1)$ and
$\phi(\lambda;M,s-1)=\Omega(\lambda;M-s+2m,s-2m)$.
Inserting this in (\ref{omla-appc-eq}) it follows proposition {\it (ii)}.

\vspace*{0.2cm}
{\sl Proof of Theorem C.1.\hspace{0.2cm}} 

According to Lemma C.2. and the definitions
of $\tilde p_{s}$ and $\tilde t_{m}(s)$,
\begin{eqnarray}
1-\tilde t_{m}(s)+\tilde p_{s}
&=&\frac{\Omega(\lambda;s-1,0)}{\Omega(\lambda;M,0)}
\left[\phi(\lambda;M,s)+\lambda
  e^{-u(s)}\Omega(\lambda;M-s,s)\right]\nonumber\\
&=&\frac{\Omega(\lambda;s-1,0)}{\Omega(\lambda;M,0)}\phi(\lambda;M,s-1)\,.
\label{onemin-appc-eq}
\end{eqnarray}
Since $\Omega(\lambda;s-1,0)=0$ for $s<2m$ and $\Omega(\lambda;M-s,s)=0$ for
$s>M+1-2m$, it follows $\tilde p_{s}=0$ for $s\notin\{2m,\ldots,M+1-2m\}$.
For $s\in \{2m,\ldots,M+1-2m\}$ on the other hand, we obtain from Lemma C.2.
and eqs.~(\ref{pllM-appc-eq},\ref{diff-appc-eq},\ref{onemin-appc-eq})
\begin{eqnarray}
0&=&\!-\!\!\log\lambda\!+\!u(l)\!+\!\log\left[\lambda
  e^{\!-\!u(l)}\frac{\Omega(\lambda;s\!-\!1,0)
\Omega(\lambda;M\!-\!l,l)}{\Omega(\lambda;M,0)}\right]\nonumber\\
&&{}\!+\!\sum_{s=l+1}^{l\!+\!2m\!-\!1}
\log\left[\frac{\Omega(\lambda;s\!-\!1,0)}
           {\Omega(\lambda;M,0)}\phi(\lambda;M,s\!-\!1)\right]
\!-\!\sum_{s=l}^{l+2m-1}
\log\left[\frac{\Omega(\lambda;s\!-\!1,0)}
         {\Omega(\lambda;M,0)}\phi(\lambda;M,s)\right]\nonumber\\
&=&\log\Omega(\lambda;M-l,l)-\log\phi(\lambda;M,l+2m-1)\nonumber\\
&=&0\,.
\end{eqnarray}

\vspace{0.3cm}
\begin{center}
{\bf ACKNOWLEDGMENTS}
\end{center}
We thank the Deutsche Forschungsgemeinschaft (SFB 513 and Heisenberg
fellowship for P.M.) for financial support.


\newpage\normalsize\noindent
\hspace*{-0.3cm}{\Large\bf Figure Captions}

\begin{itemize}
  
\item[{Fig.~1}] Occupation probability $p_\infty(l)$ of hard rod
  centers as a function of the distance $l$ from a hard wall for {\it
    (a)} three small rod lengths and a large mean occupation number
  $\bar p=0.1$, and {\it (b)} three large rod lengths and a small mean
  occupation number $\bar p=0.02$ corresponding to a continuum-like
  situation. The solid lines in {\it (a)} were drawn as a guide for
  the eye.
  
\item[{Fig.~2}] Correlation function $C(l)$ between the first possible
  position $2m$ of a rod center and another rod center that is at
  distance $l$ from the wall (see eq.~\ref{cl-eq}) )for the same
  parameters as in Figs.~1a,b. The solid lines in {\it (a)} were drawn
  as a guide for the eye.

\item[{Fig.~3}] Occupation probability $p_\infty(l)$ of rod centers as
  a function of the distance $l$ from a wall for {\it (i)} hard rods
  in the presence of a hard wall ($v=v_{\scriptscriptstyle\rm HR}$,
  $u=0$), {\it (ii)} hard rods in the presence of a soft wall
  ($v=v_{\scriptscriptstyle\rm HR}$, $u=u_0$), and {\it (iii)} rods
  with a Lennard-Jones type Takahashi interaction in the presence of a
  hard wall ($v=v_{\scriptscriptstyle\rm LJ}$, $u=0$). In {\it (a)}
  the discrete nature of the lattice is important ($m=4$, $\bar
  p=0.1$), while in {\it (b)} the data correspond to a continuum-like
  situation ($m=18$, $\bar p=0.02$). The solid lines in {\it (a)} were
  drawn as a guide for the eye.
  
\item[{Fig.~4}] Correlation function $C(l)$ between the first possible
  position $2m$ of a rod center and another rod center that is at
  distance $l$ from the wall (see eq.~\ref{cl-eq}) for the same
  parameters as in Figs.~3a,b. The solid lines in {\it (a)} were drawn
  as a guide for the eye.

\end{itemize}

\pagebreak

\begin{minipage}{15cm}
\epsfxsize=15cm
\epsffile{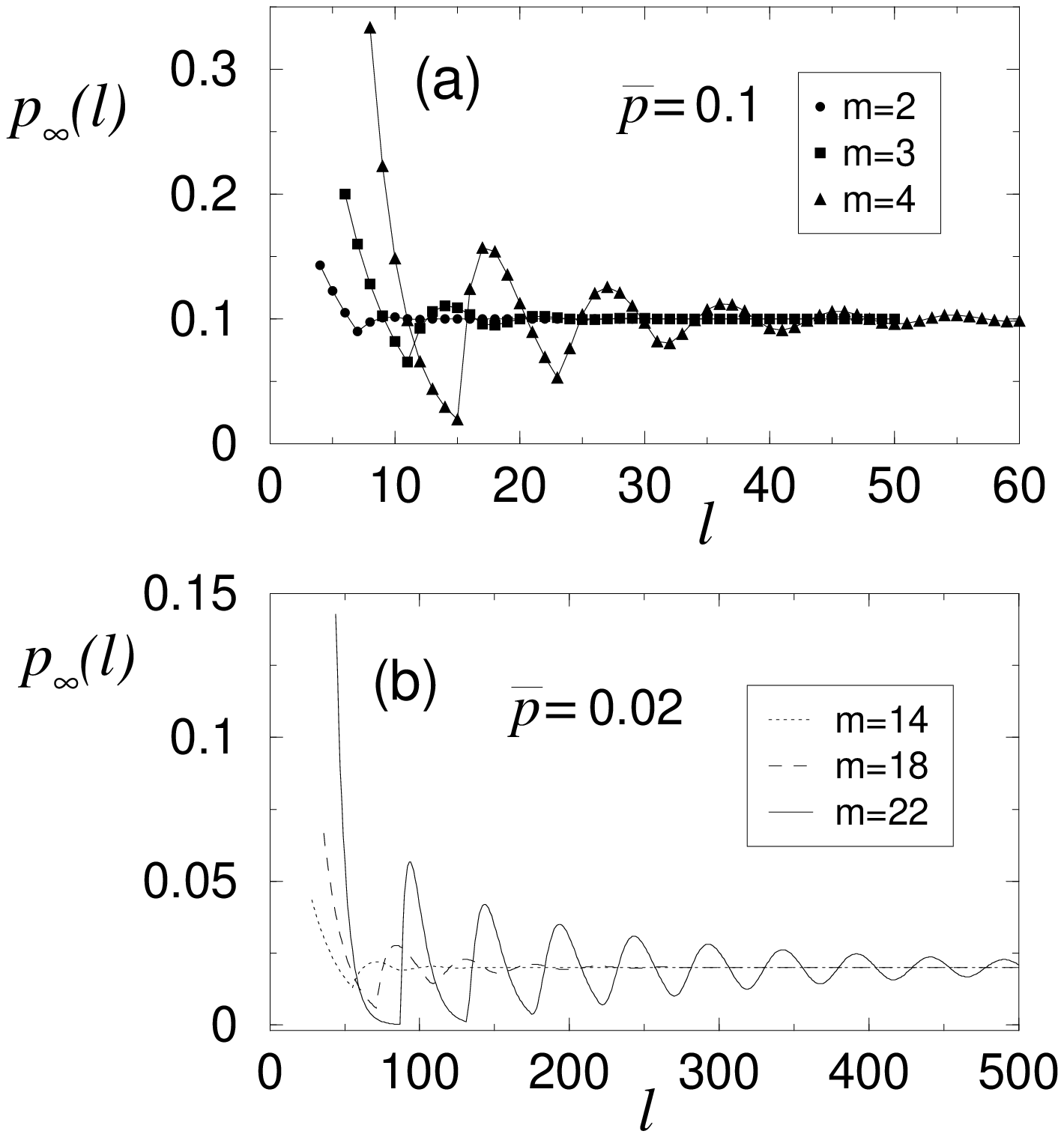}
\end{minipage}

\vspace*{0.5cm}\noindent FIG.~1.\/ Occupation probability
$p_\infty(l)$ of hard rod centers as a function of the distance $l$
from a hard wall for {\it (a)} three small rod lengths and a large
mean occupation number $\bar p=0.1$, and {\it (b)} three large rod
lengths and a small mean occupation number $\bar p=0.02$ corresponding
to a continuum-like situation. The solid lines in {\it (a)} were drawn
as a guide for the eye.

\vfill
\begin{center}
\Large (Fig.1, Buschle {\it et al.})
\end{center}

\pagebreak

\begin{minipage}{15cm}
\epsfxsize=15cm
\epsffile{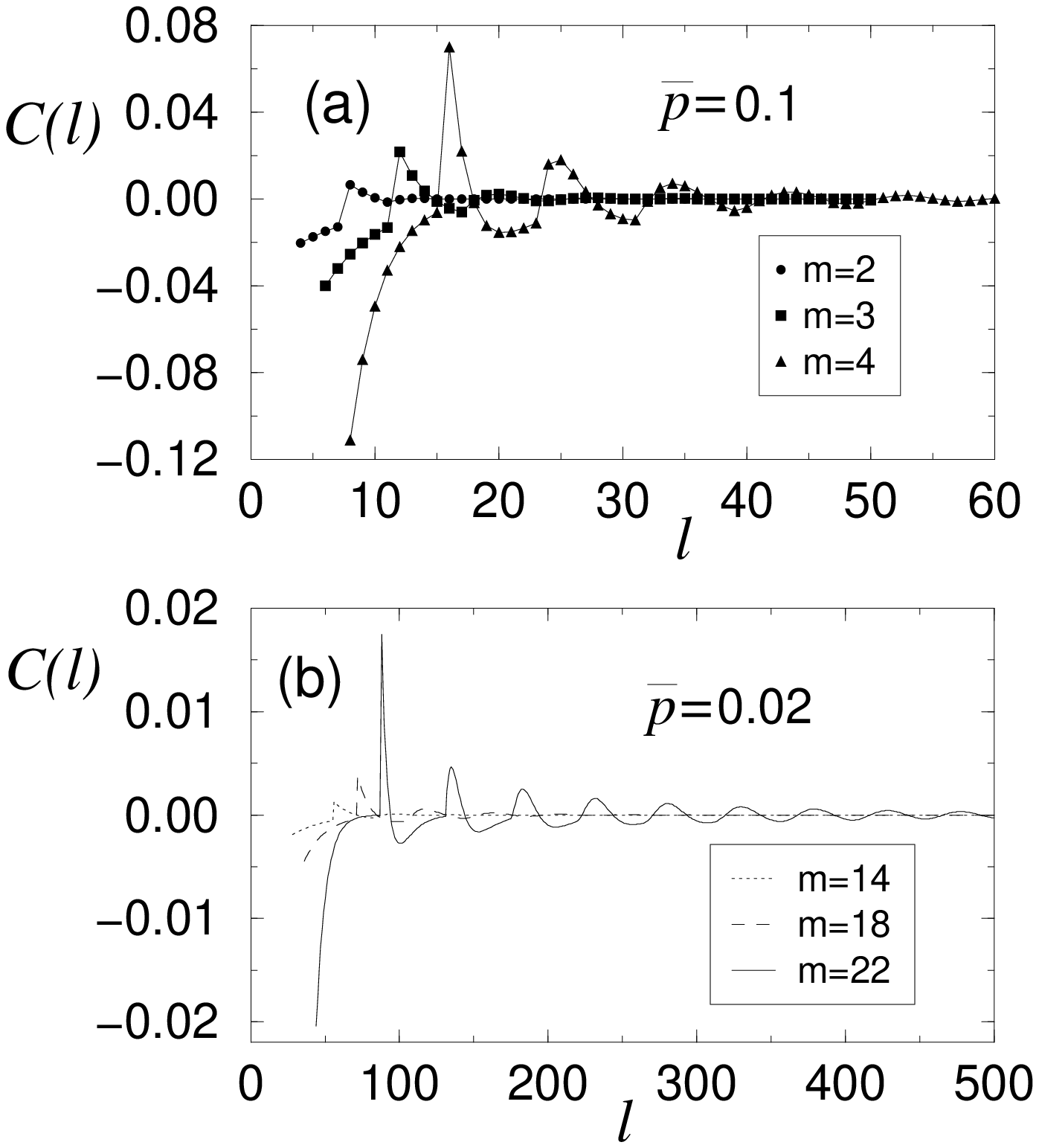}
\end{minipage}

\vspace*{0.5cm}\noindent FIG.~2.\/ Correlation function $C(l)$ between
the first possible position $2m$ of a rod center and another rod
center that is at distance $l$ from the wall (see eq.~\ref{cl-eq})
)for the same parameters as in Figs.~1a,b. The solid lines in {\it
  (a)} were drawn as a guide for the eye.

\vfill
\begin{center}
\Large (Fig.2, Buschle {\it et al.})
\end{center}

\pagebreak

\begin{minipage}{15cm}
\epsfxsize=15cm
\epsffile{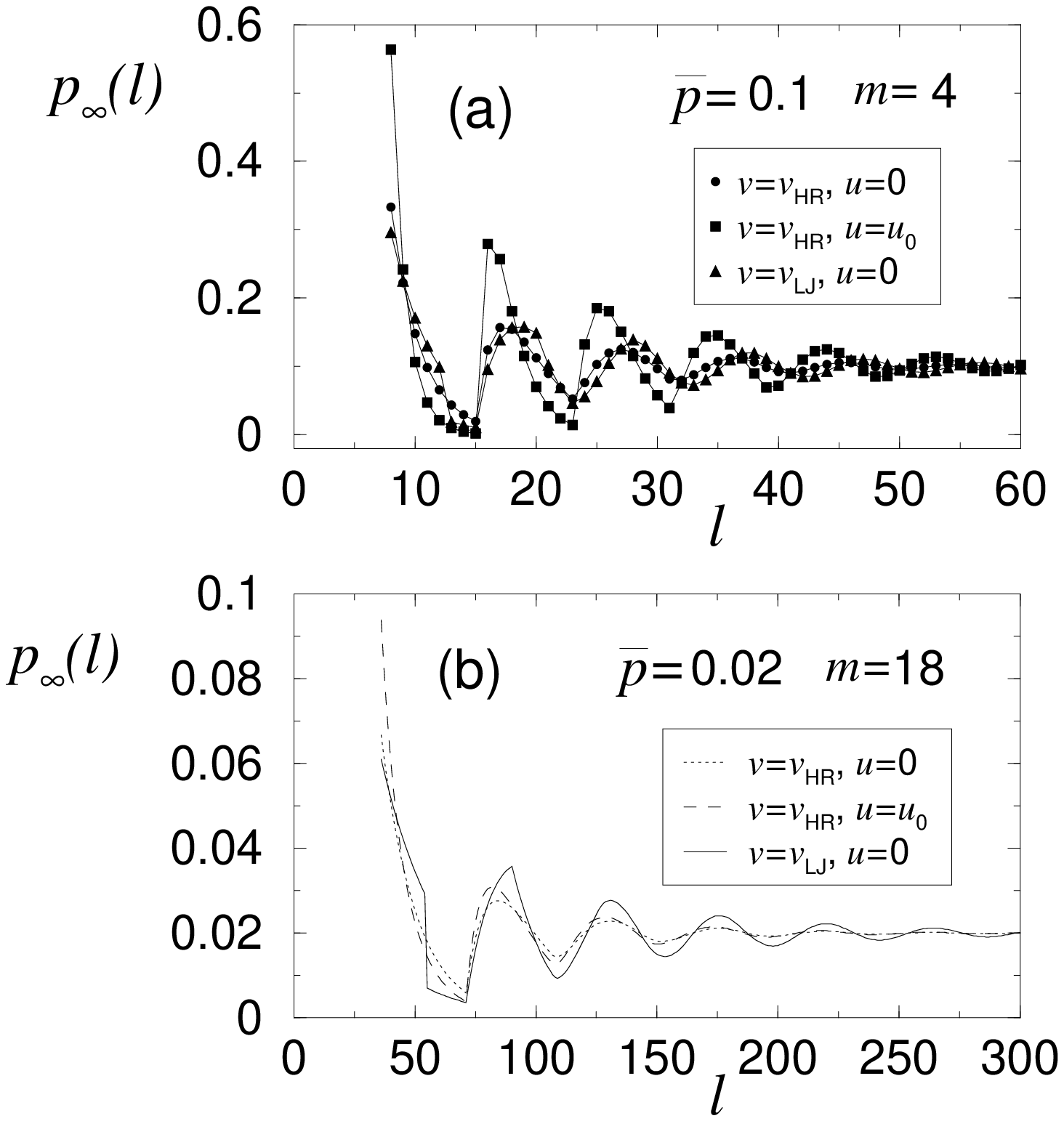}
\end{minipage}

\vspace*{0.5cm}\noindent FIG.~3.\/ Occupation probability
$p_\infty(l)$ of rod centers as a function of the distance $l$ from a
wall for {\it (i)} hard rods in the presence of a hard wall
($v=v_{\scriptscriptstyle\rm HR}$, $u=0$), {\it (ii)} hard rods in the
presence of a soft wall ($v=v_{\scriptscriptstyle\rm HR}$, $u=u_0$),
and {\it (iii)} rods with a Lennard-Jones type Takahashi interaction
in the presence of a hard wall ($v=v_{\scriptscriptstyle\rm LJ}$,
$u=0$). In {\it (a)} the discrete nature of the lattice is important
($m=4$, $\bar p=0.1$), while in {\it (b)} the data correspond to a
continuum-like situation ($m=18$, $\bar p=0.02$). The solid lines in
{\it (a)} were drawn as a guide for the eye.

\vfill\vspace*{-1.1cm}
\begin{center}
\Large (Fig.3, Buschle {\it et al.})
\end{center}

\pagebreak

\begin{minipage}{15cm}
\epsfxsize=15cm
\epsffile{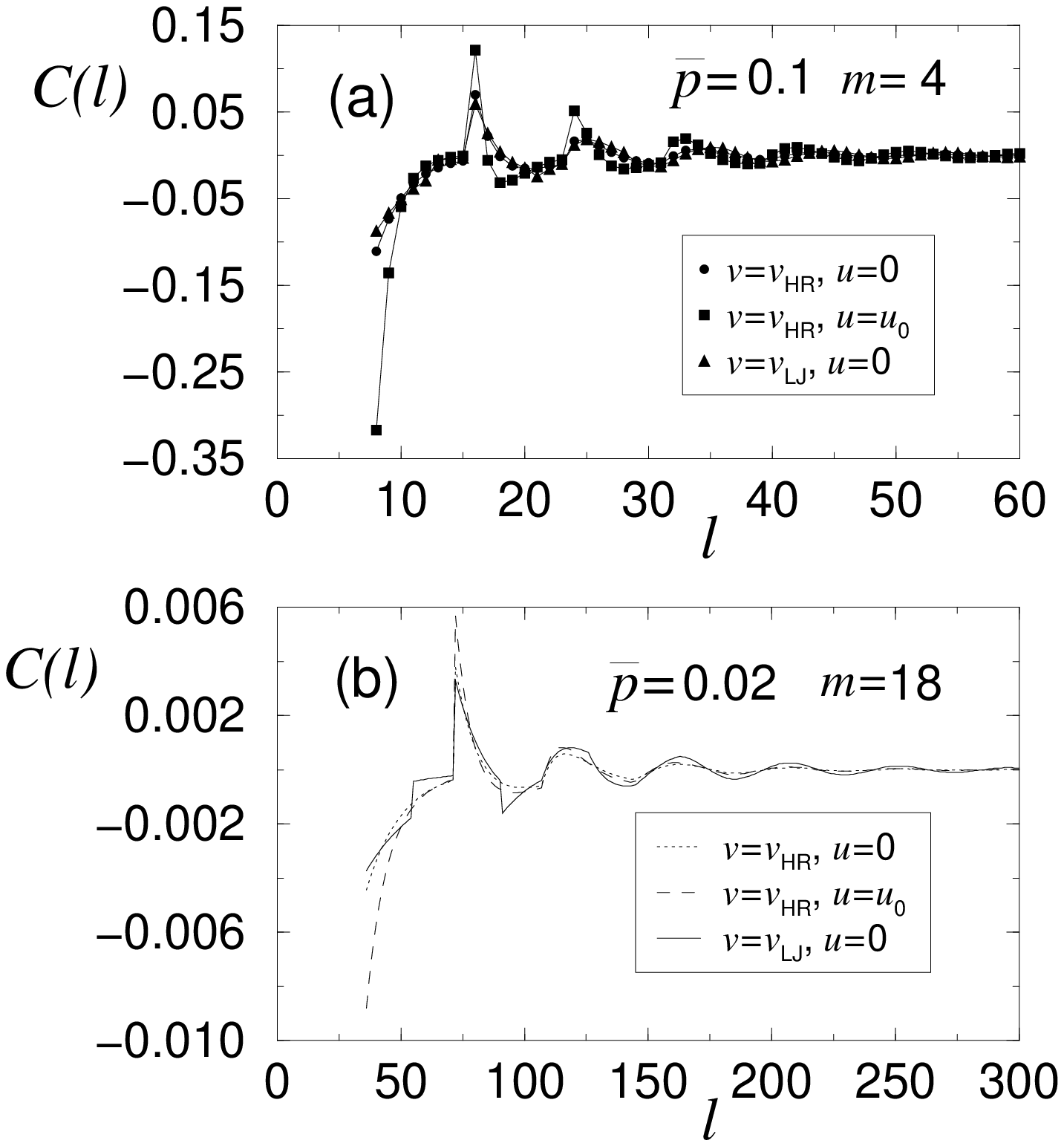}
\end{minipage}

\vspace*{0.5cm}\noindent FIG.~4.\/ Correlation function $C(l)$ between
the first possible position $2m$ of a rod center and another rod
center that is at distance $l$ from the wall (see eq.~\ref{cl-eq}) for
the same parameters as in Figs.~3a,b. The solid lines in {\it (a)}
were drawn as a guide for the eye.

\vfill
\begin{center}
\Large (Fig.4, Buschle {\it et al.})
\end{center}


\begin{thebibliography}{99}  
  
\bibitem{Croxton:1986} {\it Fluid Interfacial Phenomena}, edited by
  C.~A.~Croxton (Wiley, New York, 1986).
  
\bibitem{Chavrolin/etal:1986} {\it Liquids at Interfaces} Les Houches Summer
  School Lectures, Vol.~XLVIII, edited by J.~Chavrolin, J.~F.~Joanny, and
  J.~Zinn-Justin (Elsevier, Amsterdam, 1990).
  
\bibitem{Delamarche:1996} E.~Delamarche, B.~Michel, H.~A.~Biebuyck, and
  Christoph Gerber, Adv.~Mater. {\bf 8}, 719 (1996).
  
\bibitem{Binder:1986} K.~Binder, in {\it Phase Transitions and Critical
    Phenomena}, edited by C. Domb and J. Lebowitz, Vol. {\bf 8} (Academic
  Press, London, 1986).
  
\bibitem{Dieterich:1988} S. Dietrich, in {\it Phase Transitions and Critical
    Phenomena}, edited by C. Domb and J. Lebowitz, Vol. {\bf 12} (Academic
  Press, London, 1988) p.~1.
  
\bibitem{Schick:1990} M.~Schick, in {\it Liquids at Interfaces} Les Houches
  Summer School Lectures, Vol.~XLVIII, edited by J.~Chavrolin, J.~F.~Joanny,
  and J.~Zinn-Justin (Elsevier, Amsterdam, 1990), p.~415.
  
\bibitem{Puri/Frisch:1997} S.~Puri and H.~L.~Frisch,
J.~Phys.~Condens.~Matter {\bf 9}, 2109 (1997).

\bibitem{Evans:1992} R.~Evans, in {\it Fundamentals of Inhomogeneous
    Fluids}, edited by D.~Henderson (Marcel Dekker, New York, 1992),
  p.~85.

\bibitem{Kikuchi:1951} R.~Kikuchi, Phys.~Rev. {\bf 81}, 988 (1951).

\bibitem{Kikuchi:1966} R.~Kikuchi, Prog.~Theor.~Phys. (Kyoto)
  Suppl.~{\bf 35}, 1 (1966).
  
\bibitem{Salter/Davies:1975} S.~J.~Salter and H.~T.~Davies,
  J.~Chem.~Phys. {\bf 63}, 157 (1975).

\bibitem{Rosenfeld/etal:1996} Y. Rosenfeld, M. Schmidt, H.~L{\"o}wen,
  P.~Tarazona, J.~Phys.~Condensed Matter {\bf 8} (1996) L577.

\bibitem{Percus:1976} J.~K.~Percus, J.~Stat.~Phys. {\bf 15}, 505 (1976).
  
\bibitem{Percus:1982} J.~K.~Percus, J.~Stat.~Phys. {\bf 28}, 67 (1982).
  
\bibitem{Percus:1989} J.~K.~Percus, J.~Phys.~Condensed Matter {\bf 1},
  2911 (1989).

\bibitem{Robledo/Varea:1981} A.~Robledo and C.~Varea, J.~Stat.~Phys. {\bf 26},
  513 (1981).
  
\bibitem{Percus:1994} J.~K.~Percus, Acc.~Chem.~Rev. {\bf 27}, 8
  (1994).
  
\bibitem{Buschle:1999} J.~Buschle, {\it Diplomarbeit}, Universit\"at
Konstanz, 1999, unpublished.

\bibitem{Takahashi:1942} H.~Takahashi, Proceedings of the Physico-Mathematical
  Society of Japan (Nippon Suugaku-Buturigakkwai Kizi Tokyo) {\bf 24}: 60
  (1942). For a translation from the German see {\it Mathematical Physics in
    One Dimension}, ed. by E.~H.~Lieb and D.~C.~Mattis, (Academic Press, New
  York, 1966), p.~25.
  
\bibitem{Stanley:1986} R.~P.~Stanley, {\it Enumerative Combinatorics} Vol.~I,
  (Wadsworth \& Brooks, Belmont~/~California, 1986), p.~202.
  
\bibitem{Leff/Coopersmith:1967} H.~S.~Leff and M.~H.~Coopersmith,
  J.~Math.~Phys. {\bf 8}, 306 (1967).

\bibitem{Flicker:1968} M.~Flicker, J.~Math.~Phys. {\bf 9}, 171 (1968).
  
\bibitem{Mermin:1965} N.~D.~Mermin, Phys.~Rev. B {\bf 137}, 1441 (1965).
    
\bibitem{Lenard:1961} A.~Lenard, J.~Math.~Phys. {\bf 2}, 682 (1961), see Lemma
  3.
  
\bibitem{Feller:1968} W.~Feller, {\it An introduction to probability theory
    and its applications}, Vol.~I (Wiley \& Sons, New York, 1968), p.~330.
  
\bibitem{Good:1957} I.~J.~Good, Ann.~Math.~Stat. {\bf 28}, 861 (1957).

\bibitem{Riordan:1968} J.~Riordan, {\it Combinatorial Identities} (Wiley \&
  Sons, New York, 1968).
  
\bibitem{Goulden/Jackson:1983} I.~P.~Goulden and D.~M.~Jackson, {\it
    Combinatorial Enumerations}, (Wiley \& Sons, New York, 1983), p.~17.

\end{thebibliography}
\end{document}